\documentclass[aip,floatfix,graphicx]{revtex4-1}

\usepackage{graphicx}
\usepackage{amssymb}

\makeatletter
\newcommand*{\rom}[1]{\expandafter\@slowromancap\romannumeral #1@}
\makeatother

\draft 

\begin{document}


\title{Modification of particle-laden near-wall turbulence: effect of Stokes number} 



\author{Junghoon Lee}
\affiliation{Department of Computational Science and Engineering, Yonsei University, 50 Yonsei-ro, Seodaemun-gu, Seoul, Korea}

\author{Changhoon Lee}
\affiliation{Department of Mechanical Engineering and Department of Computational Science and Engineering, Yonsei University, 50 Yonsei-ro, Seodaemun-gu, Seoul, Korea}
\email{clee@yonsei.ac.kr}


\date{\today}

\begin{abstract}
Turbulent channel flows laden with particles are investigated using direct numerical simulation with a point-force approximation 
for small, heavy particles with a diameter smaller than the Kolmogorov length scale of the fluid.
The Stokes numbers based on the wall units considered in our study are $St^+=0.5, 5, 35$ and 125.
The main purpose of this study is 
to examine the effect of Stokes number on turbulence modification in a channel.
We found that particles with $St^+=0.5$ enhance turbulence by
increasing the occurrence of quasistreamwise vortices,
while larger-Stokes-number particles attenuate turbulence.
When $St^+=0.5$,
kinetic energy is transferred from the particles to streamwise fluid velocity fluctuations
in the high-speed regions and low-speed streaks, which may increase the instability of the low-speed streaks responsible for the birth of new quasistreamwise vortices.
On the other hand, the preferential concentration of larger-Stokes-number particles in low-speed streaks is responsible for turbulence attenuation,
and the slow response of the particles to the fluid produces feedback against the fluid velocity associated with quasistreamwise vortices.

\end{abstract}


\maketitle 

\section{INTRODUCTION}
Turbulent flows laden with particles occur frequently in nature, such as the formation of rain in clouds,
as well as in many engineering applications, such as chemical reactors. 
In these flows, dispersed particles can modify carrier turbulence
as their loading increases.

Several investigators have attempted to identify 
the factors influencing such turbulence modification (e.g. turbulence augmentation or attenuation).
For example, Gore and Crowe,\cite{Gore89} Hetsroni\cite{Hetsroni89} and Elghobashi\cite{Elghobashi91,Elghobashi94}
classified turbulence modification 
according to the ratio of particle diameter to an integral fluid length scale, particle Reynolds number and
Stokes number, respectively.
Recently, Tanaka and Eaton\cite{Tanaka08}
defined the particle momentum number by nondimensionalizing the modified Navier-Stokes equation due to the presence of particles. 

The effect of Stokes number on turbulence modification by small particles  
was systematically studied by 
Ferrante and Elghobashi\cite{Ferrante03}
and Abdelsamie and Lee\cite{Abdelsamie12}
using direct numerical simulations (DNS) of homogeneous decaying isotropic turbulence.
They considered particles smaller than the Kolmogorov length scale of the fluid.
The Stokes number range was $0.1 \leq St_K  \leq 5.0$,
where $St_K$ is the Stokes number based on the  Kolmogorov time scale.
General observations indicated that
particles with small Stokes numbers ($St_K<1$)
increase the turbulence kinetic energy, enstrophy and viscous dissipation rate,
while particles with large Stokes numbers ($St_K>1$) decrease them. 
However, particles with a $St_K\approx1$ had relatively less influence on the fluid.
Abdelsamie and Lee\cite{Abdelsamie12} further showed that
acceleration is modified 
in a similar manner to turbulence kinetic energy.
Both studies demonstrated that
Stokes number is a suitable parameter for the evaluation of turbulence modification by small particles.
Recently, Lucci \emph{et al.}\cite{Lucci11} demonstrated that this is not true for larger particles, however.
The present study considers a wide range of Stokes numbers
to investigate turbulence modification by small particles,
similar to the studies by Ferrante and Elghobashi\cite{Ferrante03} and Abdelsamie and Lee,\cite{Abdelsamie12}
but in near-wall turbulence. 

In turbulent channel flows laden with particles, many researchers have investigated the behavior of particles and turbulence, either experimentally or numerically.
Rogers and Eaton\cite{Rogers91} experimentally investigated the modification of a vertical turbulent boundary layer in air by particles using copper particles of 70 $\mu m$ in diameter. 
The Stokes numbers based on the eddy turnover time scale were of order unity.
Fluid turbulence was attenuated by the particles.
Kulick \emph{et al.}\cite{Kulick94} experimentally investigated 
solid/air two-phase turbulent flows in a vertical channel utilizing glass spheres with diameters of 50 and 90 $\mu m$,
and copper spheres with a diameter of 70 $\mu m$.
The Stokes numbers considered included $St_K\approx8, 19$ and 41 at the channel centerline,
and, correspondingly, $St^+\approx300, 700$ and 1500,
where $St^+$ is the Stokes number based on wall units (i.e. the friction velocity and the kinematic viscosity).
The particles attenuated the turbulence intensities and
the degree of turbulence attenuation increased with increasing Stokes number.
Yamamoto \emph{et al.}\cite{Yamamoto01} 
performed large-eddy simulation (LES) using the same conditions as Kulick \emph{et al.}\cite{Kulick94}
The degree of turbulence attenuation by the simulated particles
was comparable with the experimental results of Kulick \emph{et al.}\cite{Kulick94}
only for the small Stokes number case. 
Li \emph{et al.}\cite{Li01}
performed DNS of turbulent channel flow under similar conditions to Kulick \emph{et al.}\cite{Kulick94} but at a lower Reynolds number.
The Stokes numbers of glass particles with diameters of 21 and 39 $\mu m$ in a vertical channel in air were $St^+=59$ and 192, respectively.
The turbulent fluctuations were less affected by particles with larger Stokes numbers. 
Mito and Hanratty\cite{Mito06} carried out DNS of turbulent channel flow in zero-gravity. 
They observed significant turbulence attenuation by heavy particles with $St^+=200$. 
Dritselis and Vlachos\cite{Dritselis08,Dritselis11}  
considered copper particles with diameters of 13, 21, 41 and 59 $\mu m$ in air using DNS of turbulent channel flow.
The Stokes numbers were
$St_K\approx1, 3, 10$ and 21 at the channel centerline,
and, correspondingly, $St^+=10, 25, 100$ and 200.
They also considered cases with and without vertical gravity.
In both cases, the particles attenuated fluid momentum and vorticity related to coherent vortical structures.
The attenuation was more pronounced at small Stokes numbers. 
Zhao \emph{et al.}\cite{Zhao10,Zhao13} and Zhao and Andersson\cite{Zhao11}
used particles whose $St^+=1, 5, 30$ and 50 
in their DNS of turbulent channel flows.
In their simulations, the effect of gravity was neglected, however.
Particles whose $St^+=1$ hardly modified the turbulence.
As the Stokes number increased, 
fluid velocity fluctuations in the wall-normal and spanwise directions and 
Reynolds shear stress were increasingly damped.
However, streamwise fluid velocity fluctuations were enhanced by
particles whose $St^+=30$ and 50.
 Rashidi \emph{et al.}\cite{Rashidi90} and Pan and Banerjee\cite{Pan96} studied turbulence modification by near-neutral-density particles in horizontal open channel flows, performing experiments and DNS, respectively.
In their studies, the particles accumulated in low-speed streaks.
Small particles with small Stokes numbers ($St^+<1$) suppressed turbulence intensity and Reynolds stress, while larger particles enhanced them.
Unlike isotropic turbulence, the enhancement of turbulence by small particles with small Stokes numbers has not been reported previously in the study of channel turbulence. In previous studies dealing with particles with small Stokes numbers,\cite{Rashidi90, Pan96, Zhao11} turbulence was hardly modified by particles with $St^+=1$ or turbulence was suppressed by particles with $St^+<1$.

The objective of the present study is 
to investigate the effect of Stokes number on turbulence modification in a channel, using DNS, with an emphasis on small Stokes number to resolve the inconsistency between isotropic turbulence and channel turbulence.
Small particles with a diameter smaller than the Kolmogorov length scale are addressed in this study,
and thus the particle reaction with the fluid is implemented using a point-force approximation.
Four different cases classified by $St_K\approx0.037, 0.367, 2.572$ and 9.185 at the channel centerline,
and, correspondingly, $St^+=0.5, 5, 35$ and 125, were simulated.
While the particle diameter remains constant, 
the ratio of particle to fluid density varies from about 35 to 8650 for different Stokes numbers.
The effect of gravitational settling is eliminated to focus on the interaction of particles with coherent turbulence structures.
Our study reveals that
near-wall turbulence is modified by small, heavy particles 
in different ways depending on the Stokes number.
In particular, we find that
particles with a small Stokes number,
wherein the particle follows the fluid velocity until a slip occurs between the two phases
due to the small but finite inertia of the particle, 
lead to an increased occurrence of near-wall quasistreamwise vortices,
which, in turn, augments turbulence statistics in the wall region.
To the best of our knowledge, this has not been previously reported.
Also, we examine how the physical mechanisms for turbulence modification differ for cases of larger Stokes numbers.

In the following section, we describe the numerical procedures used in this study;
the numerical details of channel flow simulation and Lagrangian particle tracking are described in Sec. \rom{2} A and Sec. \rom{2} B, respectively, 
and simulation parameters are presented in Sec. \rom{2} C.
The results are presented and discussed in Sec. \rom{3};
the effects of Stokes number on turbulence statistics are presented in Sec. \rom{3} A, 
the effects of Stokes number on near-wall turbulence structures are illustrated in Sec. \rom{3} B
and turbulence modification mechanisms are examined in Sec. \rom{3} C,
focusing on the interactions 
between particles and near-wall turbulence structures 
such as quasistreamwise vortices and low-speed streaks.
Finally, conclusions are presented in Sec. \rom{4}.

\section{NUMERICAL PROCEDURES}
\subsection{Channel flow simulation}
The governing equations for incompressible flow laden with particles can be given by
\begin{equation}
\frac{Du_{i}}{Dt}=-\frac{1}{\rho}\frac{\partial{p}}{\partial{x_i}}
+\nu\frac{{\partial}^2{u_i}}{\partial{x_j}\partial{x_j}}
+{f_i},
\end{equation}
\begin{equation}
\frac{\partial{u_i}}{\partial{x_i}}=0,
\end{equation}
in which
$t$ is time, $u_i$ is the fluid velocity in the $x_i$ direction, and $x_1$, $x_2$ and $x_3$ indicate streamwise ($x$), wall-normal ($y$) and spanwise ($z$) directions, respectively.
Here, $\rho$, $p$ and $\nu$ are fluid density, pressure and kinematic viscosity, respectively,
and $f_i$ is the effect of particles on the fluid in the $x_i$ direction.
The particles considered are smaller than the Kolmogorov length scale of the fluid.
Thus, a point-force approximation was employed in implementing $f_i$ in the present code, as
\begin{equation}
f_i=-\frac{1}{m_f}\sum_{k=1}^{N_p}(D_{i})_k,
\end{equation}
in which 
$m_f$ is the fluid mass of a computational cell including any given grid point,
$(D_i)_k$ is the hydrodynamic drag force acting on the $k$-th particle in the $x_i$ direction 
and the summation operator is taken over $N_p$ particles contained within the cell.

DNS of turbulent channel flow was performed via a pseudo-spectral method.
The Chebyshev-tau method in the $y$ direction and 
the dealiased Fourier expansion in the $x$ and $z$ directions were used.
Time advancing was performed
using the Crank-Nicolson scheme for the viscous term, and 
the third-order Runge-Kutta scheme was used for the nonlinear term.
For the homogeneous directions,
periodic boundary conditions were applied. 
At the walls, $u_i=0$ according to no-slip and impermeability conditions.

\subsection{Lagrangian particle tracking}
The particles considered are small, undeformable spheres.
In this study, the ratio of particle density, $\rho_p$, to fluid density is larger than unity.
Thus, the Stokes drag is most significant,
while other forces including the pressure gradient, added mass and Basset forces can be neglected.\cite{Armenio01} 
Gravity has also been neglected
to eliminate the effect of gravitational settling on turbulence modification.
Furthermore, it is expected that in zero-gravity, the effect of lift on particles will not be significant.\cite{Arcen06} 
Therefore, the particle equation of motion can be established considering only Stokes drag as follows:
\begin{equation}
m_p\frac{dv_i}{dt}=D_i=m_p\frac{\gamma}{\tau_p}\left(\tilde{u}_i-v_i\right),
\end{equation}
\begin{equation}
\frac{dq_i}{dt}=v_i,
\end{equation}
where 
$m_p$ is the particle mass,
$\tau_p=d_p^2\rho_p/(18\rho\nu)$ ($d_p$ is particle diameter) is
the particle response time scale,
$v_i$ and $\tilde{u}_i$ are
the particle velocity 
and the fluid velocity at the particle position $(q_1(t),q_2(t),q_3(t),t)$, respectively, in the $x_i$ direction.
The coefficient $\gamma$ indicates the nonlinear drag correction factor
accounting for situations where particle Reynolds number, $Re_p$, which is 
based on the slip velocity between the two phases and particle diameter, is large,
\begin{equation}
\gamma=1+0.15Re_p^{0.687}.
\end{equation}

In order to obtain $\tilde{u}_i$, 
the four-point Hermite interpolation scheme in the $x$ and $z$ directions and
the fifth-order Lagrange polynomial interpolation in the $y$ direction were used.\cite{Choi2004}
Time advancement for equations (4) and (5) was carried out 
using the third-order Runge-Kutta scheme. 
We use the flow data of fully developed channel turbulence as initial flow fields at $t=0$,
and at that time particles are homogeneously scattered over the computational domain,
with velocities identical to the interpolated fluid velocity at their position.
For particles moving outside the computational domain, periodic boundary conditions are applied in the homogeneous directions.
Particle-wall interaction is assumed to be elastic collision.

\begin{table}
\caption{Particle parameters.
$St_K$ and $St^+$ are the particle Stokes numbers 
based on the Kolmogorov time scale and wall units, respectively, of the particle-free flow, 
$\rho_p/\rho$ is the particle-to-fluid density ratio,
$N$ is the number of particles, and
$\phi_v$ is the volume fraction.
The subscript $max$ and $min$ indicate, respectively, the maximum and minimum values of the quantity.
$St^+=0.0$ hereafter indicates the particle-free case.
For the particle-laden cases,
particle mass fraction and diameter are fixed at
$\phi_m=0.3$ and $d_p=0.00283\delta$ $(\approx0.51\nu/u_\tau^*)$, respectively. 
}
\begin{tabular*}{\hsize}{@{\extracolsep{\fill}}rrrrrc}

\hline
\hline

\multicolumn1c{$St^+$}&
\multicolumn1c{$St_{K,max}$}&
\multicolumn1c{$St_{K,min}$}&
\multicolumn1c{$\rho_p/\rho$}&
\multicolumn1c{$N$}&
\multicolumn1c{$\phi_v$}
\cr

\hline

0.0 & 0.0    & 0.0    &   0.0  &        0 &  0.0                \cr

0.5 & 0.216  & 0.037  &     35 & 76640077 &  $8.7\times10^{-3}$ \cr    

5   & 2.162  & 0.367  &    346 &  7664008 &  $8.7\times10^{-4}$ \cr

35  & 15.137 & 2.572  &   2422 &  1094858 &  $1.2\times10^{-4}$ \cr

125 & 54.060 & 9.185  &   8650 &   306560 &  $3.0\times10^{-5}$ \cr
 
\hline
\hline
\end{tabular*}
\end{table}

\subsection{Simulation parameters}
When $f_i=0$ in equation (1) (i.e. particles are absent), 
the friction Reynolds number is $Re_\tau^*={u_\tau^*\delta}/{\nu}=180$ in the present simulation.
Here, $u_\tau$ and $\delta$ are the friction velocity and the channel half width, respectively,
and the superscript asterisk ($^*$) implies quantities of particle-free flow.
The same mean pressure gradient drives the flow in the streamwise direction for all cases as 
\begin{equation}
-\frac{\tau_w^*}{\delta},
\end{equation}
where $\tau_w^*=\rho u_\tau^{*^2}$ is the wall shear stress of the particle-free flow.
The channel domain in the $x, y$ and $z$ directions is $4\pi\delta\times2\delta\times(4/3)\pi\delta$ 
and the number of grids in the corresponding directions is $128\times129\times128$.
For time advancing,
we used the time step $\Delta t=0.0006\delta/u_\tau^*$ $(\approx0.1\nu/u_\tau^{*^2})$,
which is smaller than a quarter of the shortest particle response time considered.

The particles considered are classified into four different cases 
according to Stokes number.
The Stokes numbers, $St^+$, based on wall units of the particle-free flow are $St^+=0.5, 5, 35$ and 125,
and, correspondingly, the Stokes numbers, $St_K$, based on the Kolmogorov time scale of  particle-free flow at the channel centerline
are $St_{K,min}=0.037, 0.367, 2.572$ and 9.185 
and $St_{K,max}=0.216, 2.162, 15.137$ and 54.06 at the wall.
In all cases considered, the particle diameter is $d_p=0.00283\delta$ $(\approx0.51\nu/u_\tau^*)$, 
which is always smaller than the Kolmogorov length scale of the flows simulated.
Accordingly, the density ratios are $\rho_p/\rho=35, 346, 2422$ and 8650 
for the $St^+=0.5, 5, 35$ and 125 cases, respectively.
The same particle mass fraction of $\phi_m=0.3$ gives rise to 
76,640,077, 7,664,008, 1,094,858 and 306,560 real particles for the $St^+=0.5, 5, 35$ and 125 cases, respectively.
The current simulations consider dilute suspensions such that the particle volume fraction $\phi_v$ is below $10^{-3}$,
except for the case of $St^+=0.5$.
Although $\phi_v>10^{-3}$ in the case of $St^+=0.5$, the volume fraction is still on the order of $10^{-3}$ and
the particles are expected to infrequently collide with one another due to their small inertia.
Nevertheless, particle-particle collisions can be important in local regions where particles cluster.\cite{Yamamoto01,Zhao13}
In this study, however, the effect of particle-particle collisions is not taken into account, 
in order to pinpoint the interaction between particles and turbulence and facilitate explanation of this phenomenon, as other studies have done.\cite{Richter13,Zhao13} 
Detailed parameters of the particles are presented in Table \rom{1}.

\section{RESULTS AND DISCUSSION}

In near-wall turbulence, there is a tendency for particles to preferentially accumulate in the wall region
due to their interaction with large coherent turbulence structures such as
near-wall quasistreamwise vortices.\cite{Cleaver75,Pedinotti92,Brooke92,Marchioli02} 
This phenomenon, often called \emph{preferential concentration},
is maximized when particle response time and a certain characteristic fluid time-scale match,
i.e. within a particular Stokes number range.\cite{Liu74,McCoy77,Marchioli07}
To quantify preferential concentration, 
Fig. 1 shows the wall-normal profiles of the mean particle number density, $\bar{n}$, at $t=125\nu/u_\tau^{*^2}$ and $500\nu/u_\tau^{*^2}$
normalized by its initial value, $\overline n_0$, at $t=0$,
where the particle number density, $n(x,y,z,t)$, is defined as the number of particles, $N_p$, 
per computational cell volume, including the grid point $(x,y,z)$ at time $t$,
and an overbar indicates a space-average done over an $(x,z)$ plane.

Peak concentrations appear within the viscous sublayer for all Stokes numbers
and the maximum particle concentration near the wall occurs for particles with $St^+=35$.   
An investigation reveals that during the entire period of simulation of $t=500\nu/u_\tau^{*^2}$, 
the peak particle concentration continues growing for all Stokes numbers; clearly particles with $St^+=0.5$ accumulate very slowly.
In our study, we present time-averaged turbulence statistics over a relatively early period
to assess the immediate effect of Stokes number on turbulence modification by particles.
The time-average starts when $t=125\nu/u_\tau^{*^2}$, which is equivalent to the longest particle response time scale in this study, after the release of particles into the flow domain,
and is carried out over a period of $t=375\nu/u_\tau^{*^2}$.
A space-average for the turbulence statistics is done over the $(x,z)$ plane.
Hereafter, time- and space-averaged quantities
are denoted by angle brackets $\langle \cdot \rangle$, which are distinguished from space-averaged quantities denoted by an overline.

\begin{figure}[t]
     \includegraphics[angle=0, width=\textwidth]{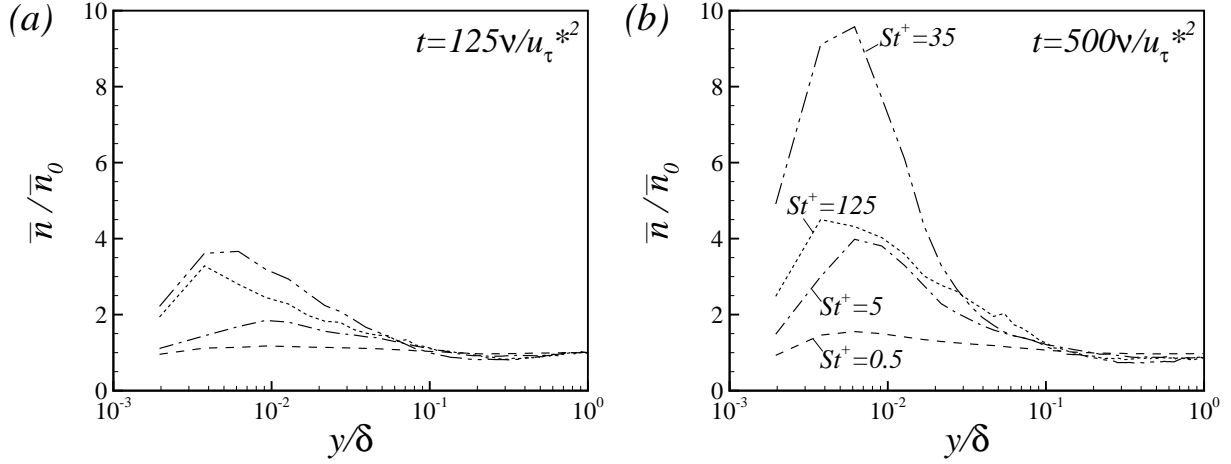}
  \caption{Wall-normal mean particle number density profiles $(a)$ at $t=125\nu/u_\tau^{*^2}$ and $(b)$ at $t=500\nu/u_\tau^{*^2}$ 
  normalized by their initial values at $t=0$.
  Dashed line, $St^+=0.5$;
  dash-dotted line, $St^+=5$;
  dash-dot-dotted line, $St^+=35$;
  dotted line, $St^+=125$.
  }
  \label{figure1}
\end{figure}

\subsection{Effects of Stokes number on turbulence statistics}

Before presenting the effects of particles, the mean momentum balance 
in the presence of particles is investigated.
The streamwise mean-momentum equation on the wall-normal interval $[0,2\delta]$ can be given by
\begin{equation}
\frac{1}{u_\tau^{*^2}}\frac{\partial \overline u_1 }{\partial t}=\frac{\partial}{\partial y}\left(\frac{\tau}{\tau_w^*}\right)
+\frac{1}{\delta},
\end{equation}
where $\overline u_i$ is the mean velocity averaged over an $(x, z)$ plane 
and $\tau$ is the total stress given by
\begin{equation}
\tau=\rho\nu\frac{\partial{\overline u_1 }}
{\partial{y}}-\rho\overline{u'_1u'_2}+\rho\int^{y'=y}_{y'=\delta}{\overline f_1 }(y') ~d y',
\end{equation}
where $u'_i$ is the fluctuating part of $u_i$, i.e. $u'_i=u_i-\overline u_i $. 
Figure 2 shows the time-averaged distribution of the total stress normalized by $\tau_w^*$ and its component terms. 
While $\tau/\tau_w^*$ for the particle-free case shows a linear profile with a slope $-1/\delta$,
$\tau/\tau_w^*$, profiles of the particle-laden flows deviate slightly from the linear profile,
indicating that these flows have not yet reached a statistically steady state due to slow accumulation of the particles towards the wall. 
Although the particle-laden flows remain transient, 
we focus on the results of this early stage of preferential accumulation
in order to assess the pure effect of the Stokes number on turbulence modification by particles.
Eventually 
the particles preferentially accumulate at the walls due to persistent interaction with near-wall quasistreamwise vortices for all Stokes numbers,
and, thus, it is difficult to identify the effect of the Stokes number on turbulence through investigation of late-time behavior.  Figure 2 also shows the level of particle stress relative to other stresses at this mass loading of particles (the mass loading for all Stokes numbers is 0.3). 
A similar result is also true for a statistically steady state
in a DNS study by Mito and Hanratty,\cite{Mito06}
where they plotted the cases of $St^+=200$ at mass loadings of approximately 0.12 and 0.49. 

In Figs. 2$(a-d)$, 
total shear stress at the wall scaled with the wall shear stress of the particle-free flow, $\tau_w^*$, 
is enhanced due to the presence of particles
compared to the particle-free case.
On average, $u_\tau$ is enhanced by about 9.7\% and 5.1\% compared to $u_\tau^*$ due to particles with $St^+=0.5$ and 5, respectively,
while it exhibits very slight enhancements for larger-Stokes-number cases (1.3\% when $St^+=35$ and 1.4\% when $St^+=125$).
Since $u_\tau$ varies, $u_\tau^*$ is used to nondimensionalize 
the variables and coordinates shown in all figures.
This provides a convenient way to
compare results for the particle-laden flows 
with the particle-free case.
Hereinafter, the superscript plus sign $^+$ denotes
quantities normalized by $u_\tau^*$ and $\nu$.

\begin{figure}[t]
     \includegraphics[angle=0, width=\textwidth]{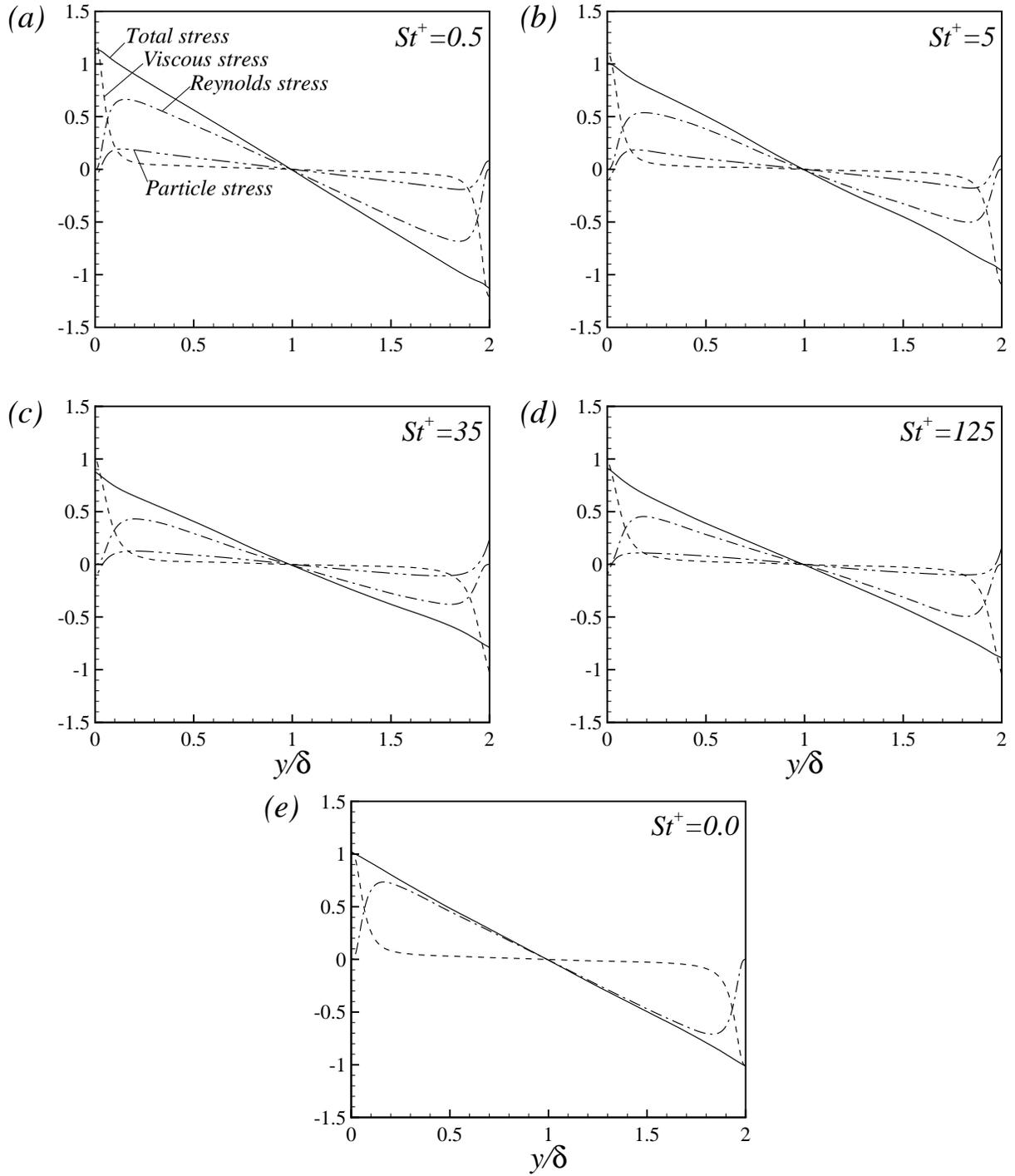}
  \caption{Total stress normalized by $\tau_w^*(=\rho u_\tau^{*^2})$, $\tau/\tau_w^*$,
  and its component terms.
  Solid line, total stress;
  dashed line, viscous shear stress;
  dash-dotted line, Reynolds shear stress;
  dash-dot-dotted line, particle stress. 
  ($a$) $St^+=0.5$; ($b$) $St^+=5$; ($c$) $St^+=35$; ($d$) $St^+=125$; ($e$) $St^+=0.0$.
  }
  \label{figure2}
\end{figure}

\begin{figure}[t]
     \includegraphics[angle=0, width=\textwidth]{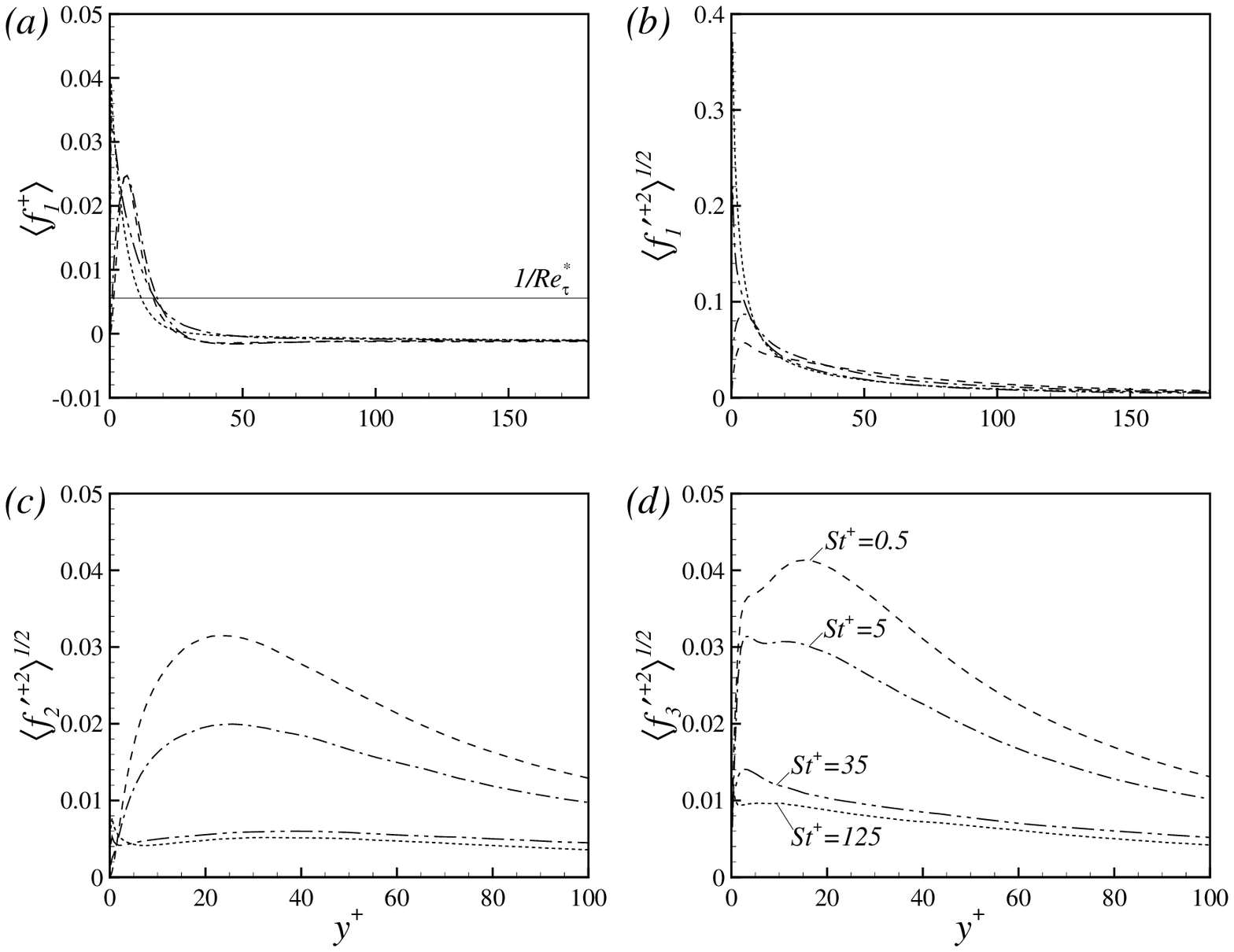}
  \caption{$(a)$ Streamwise mean and $(b)$ streamwise, $(c)$ wall-normal and $(d)$ spanwise r.m.s. particle feedback forces 
  normalized by $u_\tau^*$ and $\nu$.
  In $(a)$, mean pressure force normalized by $u_\tau^*$ and $\nu$, $1/Re_\tau^*$ is also plotted for comparison (solid line).
  Dashed line, $St^+=0.5$;
  dash-dotted line, $St^+=5$;
  dash-dot-dotted line, $St^+=35$;
  dotted line, $St^+=125$.
  }
  \label{figure3}
\end{figure}

Figure 3 presents the distribution of particle feedback forces indicating the dominant region of influence based on preferential concentration.
The streamwise mean particle force, $\langle f_1^+\rangle$,
and root-mean-square (r.m.s.) force, $\langle f_{i}'^{+^2} \rangle^{1/2}$,
where $f_{i}'$ is the fluctuating part of $f_{i}$, i.e. $f_{i}'=f_{i}-\overline f_i$, are depicted in Fig. 3.
This figure also shows the magnitude of the streamwise mean feedback force, $\langle f_1^+\rangle$, at this mass loading relative to the external mean pressure force, $1/Re_\tau^*$, both normalized by $u_\tau^*$ and $\nu$.
For all Stokes numbers, $\langle f_1^+\rangle$ has a positive peak, which is larger than the mean pressure force, near the wall.
The positive peak is located at $y^+\approx7$ when $St^+=0.5$ and 5.
As the Stokes number increases further, the positive peak increases. 
The peak is found almost at the wall for cases in which $St^+=35$ and 125.
This is qualitatively consistent with the previous result for $St^+=200$.\cite{Mito06} 
For all cases considered, 
the peak location of $\langle f_{1}'^{+^2} \rangle^{1/2}$ is almost the same as that of $\langle f_1^+\rangle$,
and its magnitude is at least two times greater than the mean value
throughout the channel width. 
This is also true for the wall-normal component, i.e. $\langle f_{2}'^{+^2} \rangle^{1/2}\gg\langle f_2^+\rangle$ (not shown here).
$\langle f_{2}'^{+^2} \rangle^{1/2}$ has a peak in the buffer layer when $St^+=0.5$ and 5.
On the other hand, 
when $St^+=35$ and 125, 
$\langle f_{2}'^{+^2} \rangle^{1/2}$ has two peaks, one appearing very close to the wall and the other in the buffer layer.
This is because these particles collide with the wall more often than particles with smaller Stokes numbers ($St^+=0.5$ and 5).
The particles colliding with the wall produce large negative and positive $f_2$.
Therefore, the local peak of $\langle f_{2}'^{+^2} \rangle^{1/2}$ is observed very close to the wall.
The spanwise component of the particle feedback force is stronger than the wall-normal component for all Stokes numbers, with peaks shifted more toward the wall, as shown in Figs. 3$(c)$ and 3$(d)$.

Figure 4 illustrates changes in streamwise mean fluid velocity $\langle u_1^+ \rangle$ due to particles. 
Particles with $St^+=0.5$ increase $\langle u_1^+ \rangle$ in the viscous sublayer and in the buffer layer
and decrease it in the region of $y^+>30$ compared to the particle-free case, although the overall degree of change is quite small.
On the other hand, the opposite trends are observed for $St^+=35$ and 125,
where a slight decrease and increase in $\langle u_1^+ \rangle$ occur in the buffer layer and in the outer part, respectively.
At a statistically steady state, more significant increases in the mean fluid velocity outside the buffer layer were observed for $St^+=30, 50, 192$ and 200 in the literature.\cite{Li01,Mito06,Zhao10,Zhao13}
For the case of $St^+=5$, 
$\langle u_1^+ \rangle$ is slightly enhanced in the channel, except for the core region where $y^+>100$.

\begin{figure}[t]
     \includegraphics[angle=0, width=0.9\textwidth]{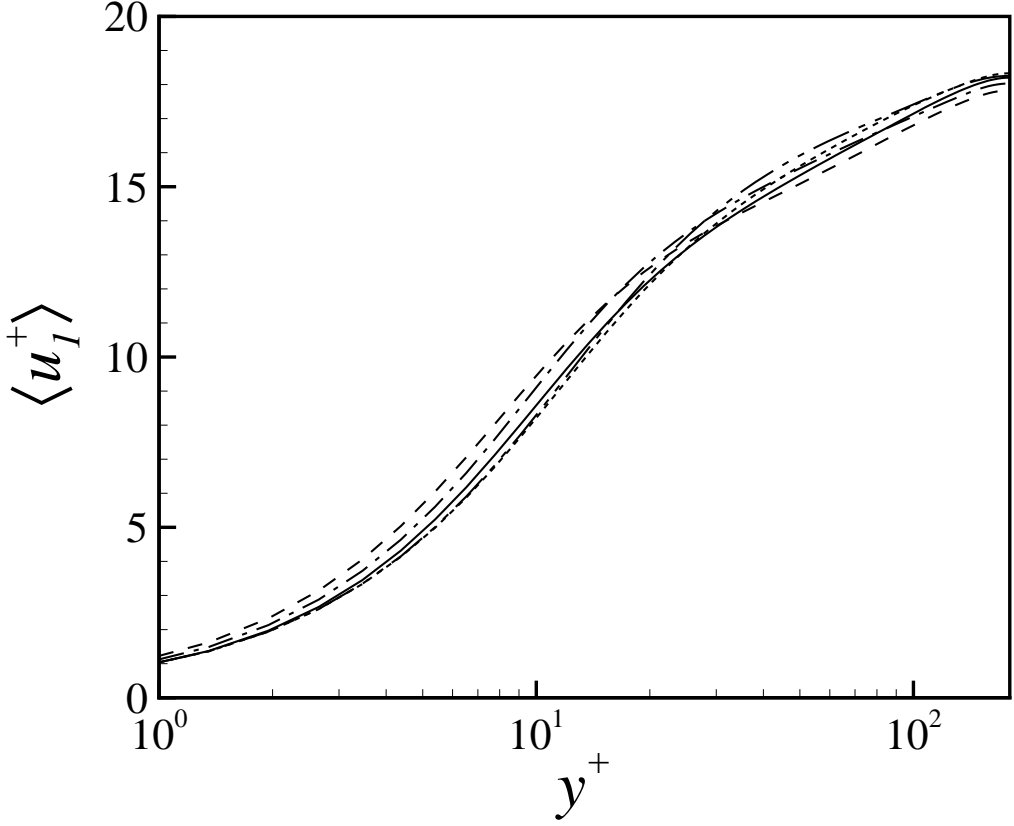}
  \caption{Mean velocities normalized by $u_\tau^*$.
  Solid line, $St^+=0.0$;
  dashed line, $St^+=0.5$;
  dash-dotted line, $St^+=5$;
  dash-dot-dotted line, $St^+=35$;
  dotted line, $St^+=125$.
  }
  \label{figure4}
\end{figure}

\begin{figure}[t]
     \includegraphics[angle=0, width=\textwidth]{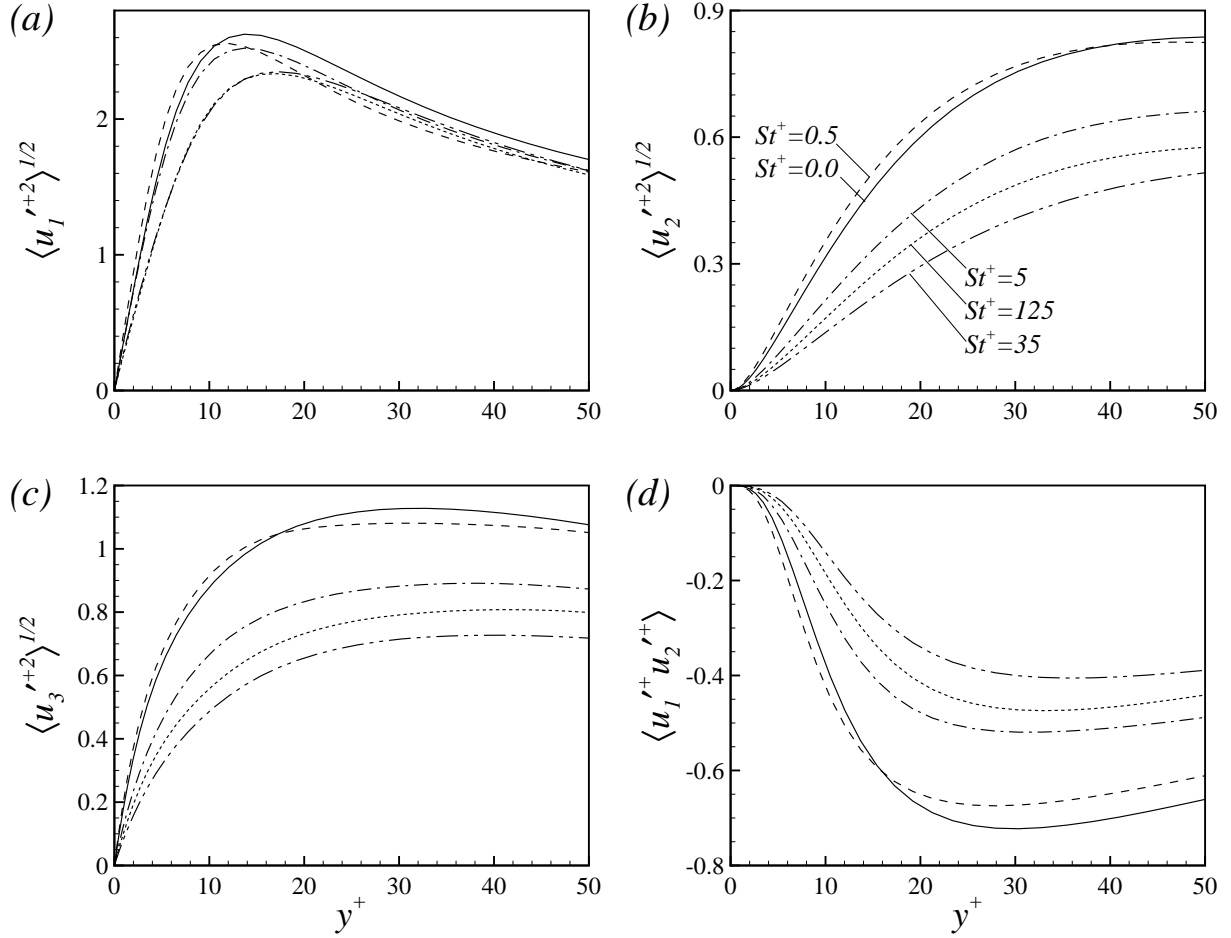}
  \caption{$(a)$ Streamwise, $(b)$ wall-normal 
  and $(c)$ spanwise turbulence intensities 
  and $(d)$ Reynolds shear stress normalized by $u_\tau^*$ in the near-wall region.
  Solid line, $St^+=0.0$;
  dashed line, $St^+=0.5$;
  dash-dotted line, $St^+=5$;
  dash-dot-dotted line, $St^+=35$;
  dotted line, $St^+=125$.
  }
  \label{figure5}
\end{figure}
\begin{figure}[t]
     \includegraphics[angle=0, width=\textwidth]{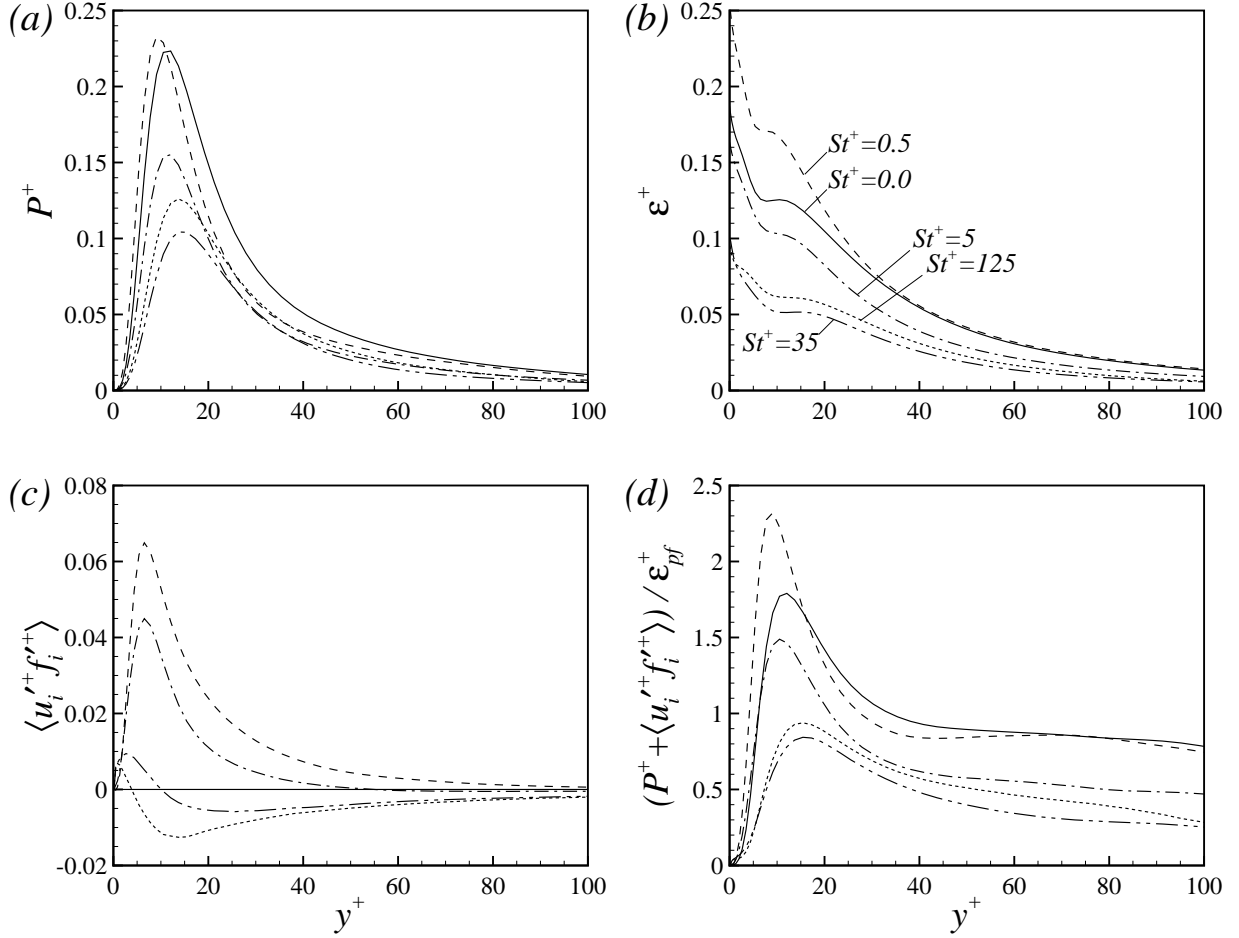}
  \caption{$(a)$ Turbulence production $P$, $(b)$ viscous dissipation $\varepsilon$, $(c)$ energy exchange between the two phases $\langle u_i'f_i' \rangle$ and $(d)$ energy transfer ratio between the sum of $P$ and $\langle u_i'f_i' \rangle$ and the dissipation of the particle-free flow $\varepsilon_{pf}$ normalized by $u_\tau^*$ and $\nu$.
  Solid line, $St^+=0.0$;
  dashed line, $St^+=0.5$;
  dash-dotted line, $St^+=5$;
  dash-dot-dotted line, $St^+=35$;
  dotted line, $St^+=125$.
  }
  \label{figure6}
\end{figure}

\begin{figure}[t]
     \includegraphics[angle=0, width=\textwidth]{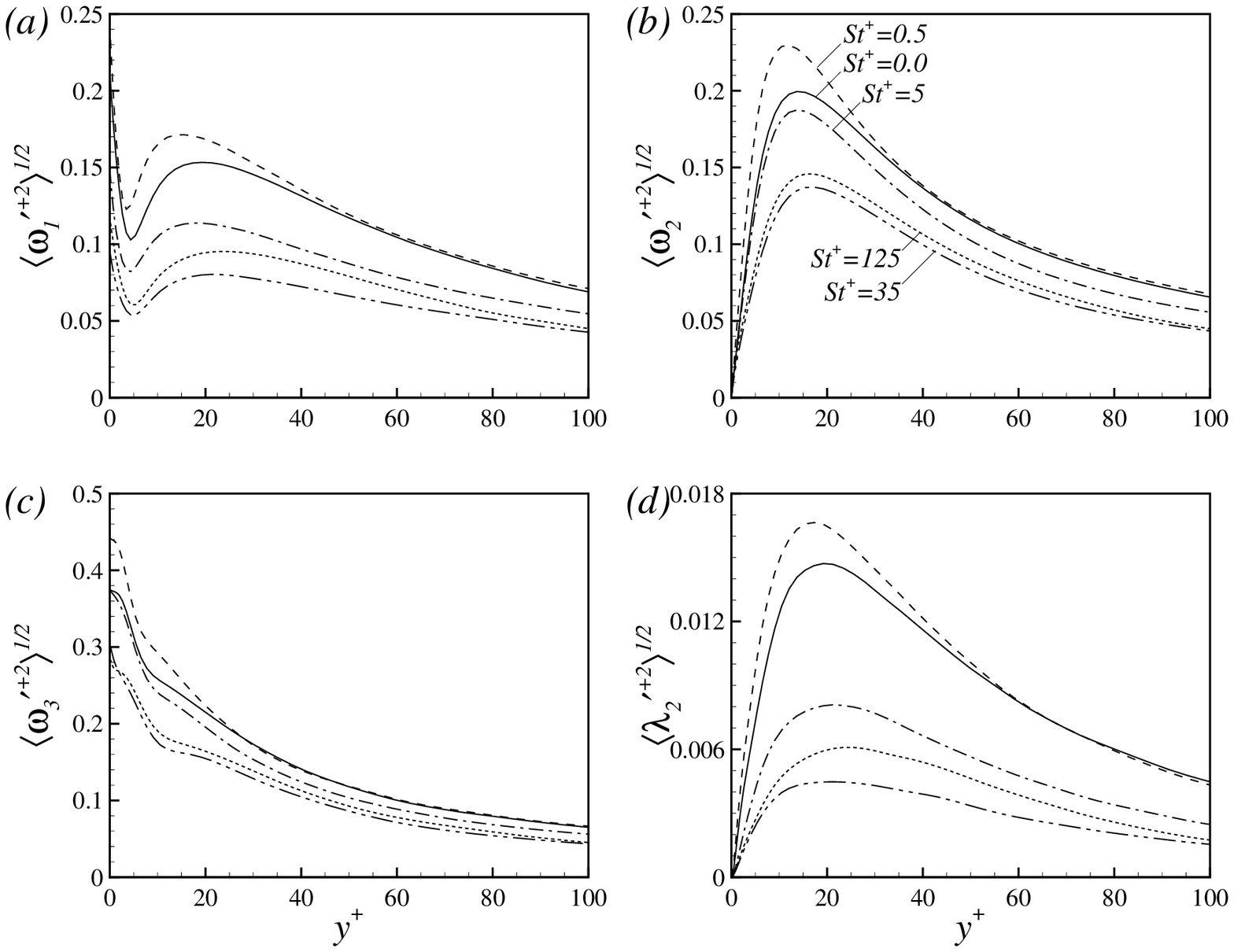}
  \caption{$(a)$ Streamwise, $(b)$ wall-normal, $(c)$ spanwise r.m.s. vorticity 
  and $(d)$ r.m.s. $\lambda_2$
  normalized by $u_\tau^*$ and $\nu$.
  Solid line, $St^+=0.0$;
  dashed line, $St^+=0.5$;
  dash-dotted line, $St^+=5$;
  dash-dot-dotted line, $St^+=35$;
  dotted line, $St^+=125$.
  }
  \label{figure7}
\end{figure}

Figure 5 shows modification of  
the turbulence intensities, $\langle u_{i}'^{+^2}\rangle^{1/2}$, 
and the Reynolds shear stress, $\langle u_{1}'^{+}u_{2}'^{+} \rangle$ by particles in the near-wall region. 
Particles with $St^+=5, 35$ and 125, as shown in Figs. 5 $(b-d)$,
suppress the wall-normal and spanwise turbulence intensities and the Reynolds stress noticeably, as compared to the particle-free case, which is consistent with the observations of the previous studies.
The most significant attenuation occurs for the case of $St^+=35$, where preferential concentration is maximized.
On the other hand, 
particles with $St^+=0.5$ 
increase turbulence slightly near the wall, and this kind of enhancement of near-wall turbulence by particles has never been observed before.

In Fig. 5$(a)$, 
no enhancement of streamwise turbulence intensity $\langle u_{1}'^{+^2} \rangle^{1/2}$ 
is observed for $St^+=5, 35$ or 125.
This does not agree with previous observations by other DNS studies for a similar Stokes number range, which found that streamwise turbulence intensity is enhanced roughly in the region outside the viscous sublayer.\cite{Li01,Dritselis08,Dritselis11,Zhao10,Zhao11,Zhao13}  
This discrepancy results from the fact that our simulations are primarily targeted at an early stage,
while the previous studies focused on a statistically steady state.
We confirmed this by simulating the cases for a longer period of time.
On the other hand, $\langle u_{1}'^{+^2}\rangle^{1/2}$
is slightly increased by particles with $St^+=0.5$ in the near-wall region
and is reduced in the region $y^+>10$.

With the particle-force term $f_i$, the mean-kinetic-energy equation for channel turbulence can be written as,
\begin{equation}
\frac{\partial{k}}{\partial t}+
\frac{\partial}{\partial y}
\left(\textstyle \frac{1}{2} \displaystyle\overline{u_{2}'u_{i}'u_{i}'}+\frac{\overline {u_{2}'p'}}{\rho}-\nu\frac{\partial}{\partial y}(k+\overline {u_{2}'u_{2}'})\right)=
-\overline{u_{1}'u_{2}'}\frac{\partial \overline u_1}{\partial y}
-2\nu\overline{s_{ij}'s_{ij}'}
+\overline{u'_if'_i},
\end{equation}
where $k=\frac{1}{2}\overline{u_{i}'u_{i}'}$ is the space-averaged turbulence kinetic energy, $p'$ is the fluctuating pressure and $s_{ij}'=\frac{1}{2}\left(\frac{\partial u_{i}'}{\partial x_j}+\frac{\partial u_{j}'}{\partial x_i}\right)$ is the fluctuating strain rate.
In this study, we focus on the changes in the terms on the right-hand side of Eq. (10) due to the presence of particles in order to further investigate their contribution to kinetic energy transfer.
Several authors have attempted to model the extra term $\overline {u'_if'_i}$ (i.e. the last term in Eq. (10)) present in particle-laden turbulent flows 
(see a recent review by Balachandar and Eaton\cite{Balachandar10}).
In this study, $\overline {u'_if'_i}$ is obtained using the results of DNS with a point-force approximation.
Figures 6$(a)$ and 6$(b)$ show the time-averaged turbulence production $P^+=-\langle{u_{1}'^+u_{2}'^+}\rangle\frac{\partial \langle u_1^+\rangle}{\partial y^+}$ 
and viscous dissipation $\varepsilon^+=2\langle{s_{ij}'^+s_{ij}'^+}\rangle$, respectively.
The presence of particles with $St^+=0.5$ leads to the enhancement of both $P^+$ and $\varepsilon^+$ in the wall region,
but $P^+$ is decreased rather than enhanced further away from the wall compared to the particle-free case.
On the other hand, larger-Stokes-number particles decrease both $P^+$ and $\varepsilon^+$,
consistent with previous results under similar conditions.\cite{Li01,Mito06,Zhao13}
The attenuation of production and dissipation is most pronounced for $St^+=35$.
The results of Fig. 6$(b)$ suggest that the presence of particles influences the length scales of turbulence,
since a smaller scale of turbulent motion results in higher dissipation, this will be shown later.
The energy exchange term, $\langle u_{i}'^{+}f_{i}'^{+} \rangle$, is shown in Fig. 6$(c)$.   
Particles with $St^+=0.5$ produce positive $\langle u_{i}'^{+}f_{i}'^{+} \rangle$
with a peak at $y^+\approx7$. 
As the Stokes number increases, $\langle u_{i}'^{+}f_{i}'^{+} \rangle$ decreases 
and becomes negative for the cases of $St^+=35$ and 125. 
However, a local positive peak occurs in the region very close to the wall for $St^+=35$ and 125. 
In order to examine 
how efficiently kinetic energy is transferred from the mean flow (i.e. of the modified flow field due to the presence of particles) and particles 
to fluid velocity fluctuations
compared to the viscous dissipation of the particle-free case,
the ratio of the sum of $P^+$ and $\langle u_{i}'^{+}f_{i}'^{+} \rangle$ to the dissipation of the particle-free flow $\varepsilon_{pf}^+$
is plotted in Fig. 6$(d)$. 
For the case of $St^+=0.5$,
$(P^++\langle u_{i}'^{+}f_{i}'^{+} \rangle)/\varepsilon_{pf}^+$ is enhanced in the wall region 
compared to the particle-free case 
and this is why particles with $St^+=0.5$ increase the turbulence intensities and the Reynolds stress near the wall in Fig. 5.
On the other hand, for larger Stokes numbers, this ratio decreases compared to the particle-free case 
and the decrease is most pronounced for the case of $St^+=35$.
This leads to the decreased turbulence intensities and Reynolds stress shown in Fig. 5.

The r.m.s. vorticity, $\langle \omega_{i}'^{+^2} \rangle^{1/2}$, and 
$\lambda_2$ distributions, $\langle \lambda_{2}'^{+^2} \rangle^{1/2}$,
are shown in Fig. 7,  
where $\omega_{i}'$ and $\lambda_{2}'$ are the fluctuating parts 
of the vorticity in the $x_i$ directions $\omega_{i}$ and $\lambda_2$, respectively.
Particles with $St^+=5, 35$ and 125 suppress $\langle \omega_{i}'^{+^2} \rangle^{1/2}$ and $\langle \lambda_{2}'^{+^2} \rangle^{1/2}$.
In particular, the attenuation of $\langle \omega_{1}'^{+^2} \rangle^{1/2}$ and $\langle \lambda_{2}'^{+^2} \rangle^{1/2}$
is most pronounced for $St^+=35$. 
The attenuation of $\omega_{i}'$ and $\lambda_{2}'$
is consistent with previous observations at similar Stokes numbers (i.e. $St^+=10, 25, 100$ and 200).\cite{Dritselis08,Dritselis11}  
On the other hand, particles with $St^+=0.5$ increase
$\langle \omega_{i}'^{+^2} \rangle^{1/2}$
and $\langle \lambda_{2}'^{+^2} \rangle^{1/2}$ near the wall.
As will be confirmed later by investigating instantaneous flow fields, 
the results of Figs. 7$(a)$ and 7$(d)$ indicate that 
particles with $St^+=0.5$ increase the occurrence of near-wall quasistreamwise vortices
while larger-Stokes-number particles decrease it, 
since $\langle \lambda_{2}'^{+^2} \rangle^{1/2}$ can be a measure of how many parts of the flow domain vortical events occupy
\cite{Jeong97,Dritselis08,Dritselis11}
and typical vortical events in near-wall turbulence are near-wall quasistreamwise vortices.
The modification of quasistreamwise vortex distribution according to Stokes number
leads to changes in turbulence production, as shown in Fig. 6$(a)$,
because near-wall quasistreamwise vortices are responsible for the generation of turbulence production.\cite{Kim87}

Figures 8 and 9 show the mean acceleration, $\langle a_i^+ \rangle$, 
and r.m.s. fluid accelerations, $\langle a_{i}'^{+^2}\rangle^{1/2}$, respectively,
where $a_i$ and $a_{i}'$ are fluid acceleration in the $x_i$ direction and its fluctuating part, respectively.
In this study, $a_i$ is obtained using the second-order time-accurate difference scheme
along the trajectory of a fluid particle in a Lagrangian frame.
To statistically evaluate fluid acceleration,
trajectories of $10^6$ randomly released fluid particles were computed 
by adopting the same interpolation schemes 
as in Lagrangian tracking for heavy particles, as described in Sec. \rom{2} B.
Fig. 8 demonstrates that the negative $\langle a_1^+ \rangle$ and positive $\langle a_2^+ \rangle$ near the wall
are enhanced due to the presence of particles with $St^+=0.5$,
while they are suppressed due to larger-Stokes-number particles,
and this suppression is most pronounced for $St^+=35$.
Since large wall-normal and spanwise fluid accelerations are mostly centripetal accelerations 
associated with quasistreamwise vortices, 
strongly converging towards the center of the vortex region,\cite{Lee04,Yeo10}
modification of the quasistreamwise vortex distribution shown in Fig. 7
naturally leads to modification of $\langle a_{2}'^{+^2}\rangle^{1/2}$ and $\langle a_{3}'^{+^2}\rangle^{1/2}$. 
Therefore, modifications of $a_{2}'$ and $a_{3}'$ are consistent with the modifications of $\omega_{1}'$ and $\lambda_{2}'$,
as shown in Figs. 9$(b,c)$.
For the streamwise component (Fig. 9$a$), 
r.m.s. values are enhanced in the viscous wall region due to the presence of particles with $St^+=0.5$,
and are attenuated by larger-Stokes-number particles. The most significant attenuation of $\langle a_{1}'^{+^2}\rangle^{1/2}$ occurs, again, for $St^+=35$.

\begin{figure}[t]
     \includegraphics[angle=0, width=\textwidth]{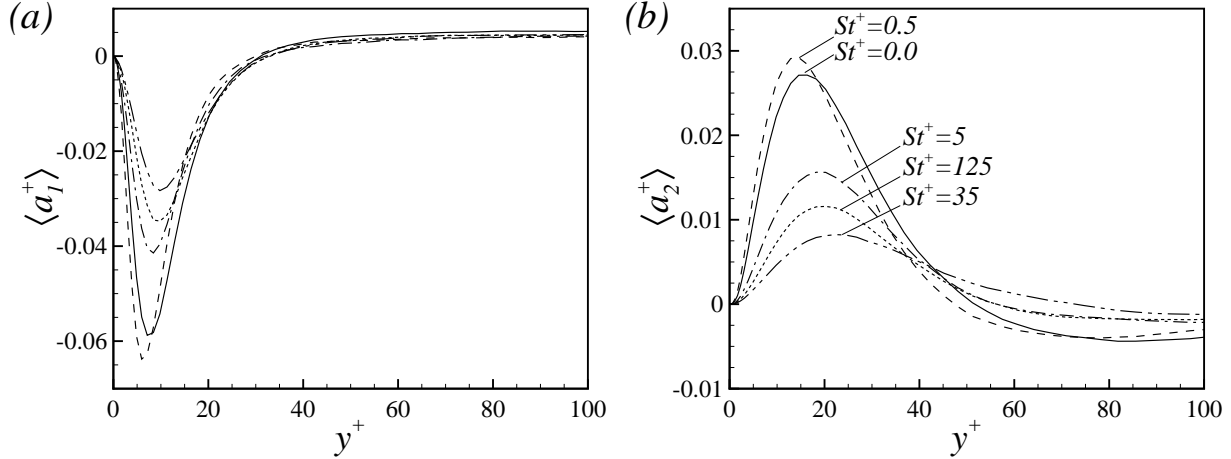}
  \caption{$(a)$ Streamwise and $(b)$ wall-normal
  mean accelerations
  normalized by $u_\tau^*$ and $\nu$. 
  Solid line, $St^+=0.0$;
  dashed line, $St^+=0.5$;
  dash-dotted line, $St^+=5$;
  dash-dot-dotted line, $St^+=35$;
  dotted line, $St^+=125$.
  }
  \label{figure8}
\end{figure}

\begin{figure}[t]
     \includegraphics[angle=0, width=\textwidth]{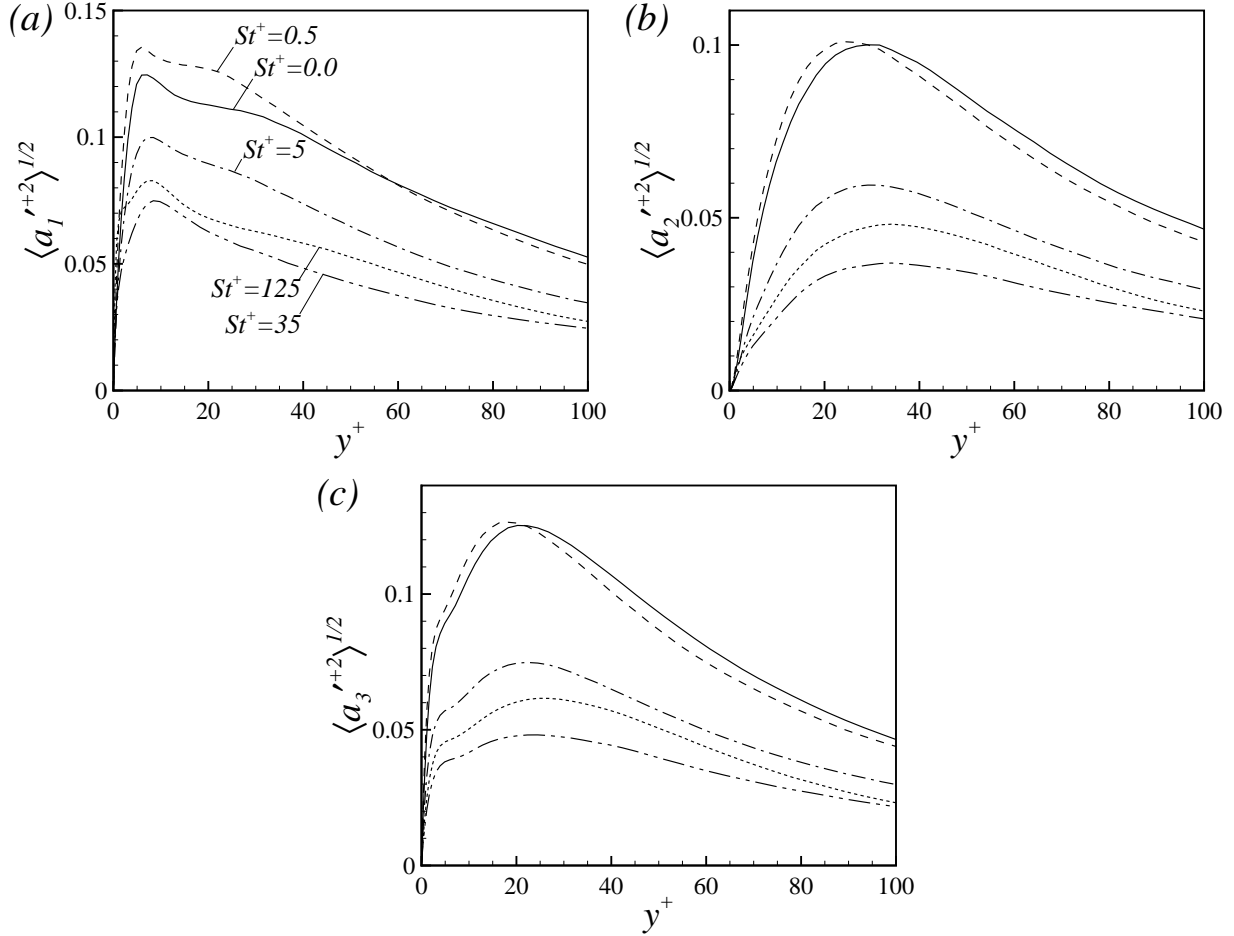}
  \caption{$(a)$ Streamwise, $(b)$ wall-normal and $(c)$ spanwise r.m.s. accelerations 
  normalized by $u_\tau^*$ and $\nu$. 
  Solid line, $St^+=0.0$;
  dashed line, $St^+=0.5$;
  dash-dotted line, $St^+=5$;
  dash-dot-dotted line, $St^+=35$;
  dotted line, $St^+=125$.
  }
  \label{figure9}
\end{figure}

\subsection{Effects of particles on near-wall turbulence structures}

Instantaneous flow fields showing near-wall turbulence structures are investigated in this section 
to gain further insight into turbulence modification by particles.
Figure 10 illustrates the instantaneous isosurfaces of $\lambda_2^+=-0.02$,
which detect vortical regions, 
and the contours of streamwise vorticity $\omega_1^+$ on the isosurfaces
to reveal modification of quasistreamwise vortices near the wall,
which are the typical coherent vortical structure of near-wall turbulence.
Compared to the particle-free case in Fig. 10$(e)$,
the number of quasistreamwise vortices increases significantly for $St^+=0.5$ (Fig. 10$a$).
On the other hand,
the attenuation of quasistreamwise vortices is observed for larger-Stokes-number cases. 
In particular, 
for $St^+=35$,
only few vortices survive (Fig. 10$c$).
The instantaneous results of Fig. 10 are consistent with changes in  
$\langle \omega_1^{+^2} \rangle^{1/2}$
and $\langle \lambda_{2}'^{+^2} \rangle^{1/2}$ depending on the Stokes number,
as shown in Fig. 7. 

A quasistreamwise vortex produces 
coherent motion of high-speed fluid towards the wall involving both $u_1'>0$ and $u_2'<0$, called \emph{sweeps},
and of low-speed fluid away from the wall including both $u_1'>0$ and $u_2'<0$, called \emph{ejections}.\cite{Robinson91a,Jeong97}
In particular, ejections are responsible for the presence of streamwise-elongated low-speed streaks.
Therefore, the different quasistreamwise vortex distributions dependent on Stokes number shown in Fig. 10
can affect the distribution of high-speed regions and low-speed streaks.
This is clearly seen in Fig. 11, where instantaneous contours of $u_1'^+$ at $y^+\approx15$ are displayed.
Li \emph{et al.}\cite{Li01} and Zhao \emph{et al.}\cite{Zhao10,Zhao13} have observed significant suppression of small scales related to streamwise velocity fluctuations 
for $St^+=192$ at a mass loading of 0.4 and for $St^+=30$ at a mass loading of 1, respectively.
Zhao \emph{et al.} further demonstrated the wider spanwise spacing between high-speed regions and low-speed streaks due to the presence of particles.
In Fig. 11, although the current mass loading is lower (i.e. $\phi_m=0.3$), 
the suppression of small scales is observed for particles with $St^+=35$ and 125, consistent with previous observations,
while finer scales are augmented by particles with $St^+=0.5$.
Relatively little influence is observed for the case of $St^+=5$.  
These modification behaviors of length scales are responsible for the modification of viscous dissipation shown in Fig. 6$(b)$.
In the case of $St^+=0.5$,
this increase of small scales and, thus, higher viscous dissipation, 
along with the increased occurrence of quasistreamwise vortices,
results in the small enhancement of the turbulence intensities and the Reynolds stress shown in Fig. 5. 
Furthermore, the increased and decreased spacing between the high- and low-speed regions according to Stokes number can lead to a respective decrease and increase in $\langle \omega_{2}'^{+^2} \rangle^{1/2}$,
since $\omega_{2}\approx\partial u_1/\partial z$ near the wall, consistent with Fig. 7$(b)$.  

\begin{figure}[t]
     \includegraphics[angle=0, width=\textwidth]{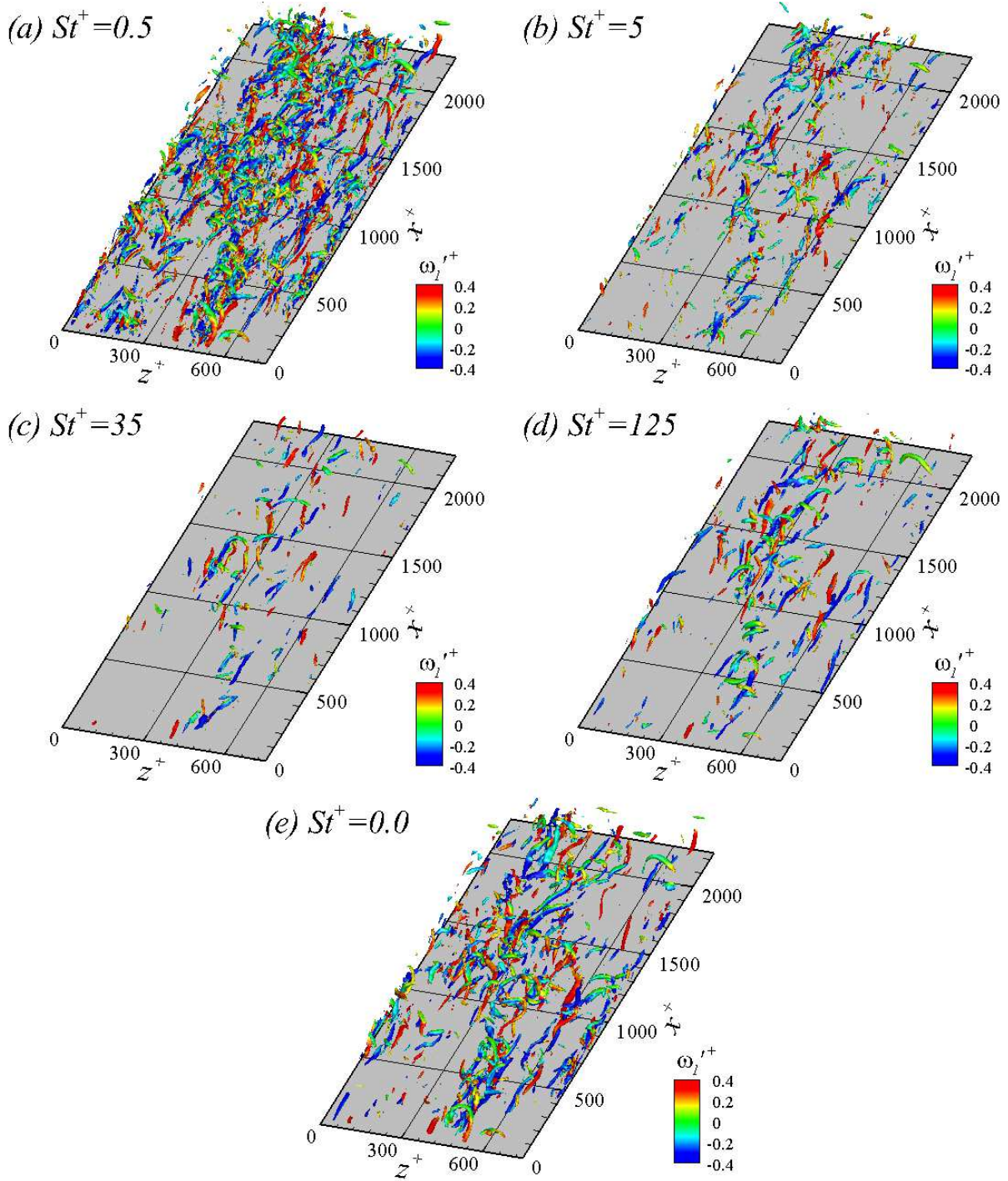}
  \caption{(Color-online) Instantaneous isosurfaces of $\lambda_2^+=-0.02$ 
  in the coordinates $(x^+,y^+,z^+)$ at $t^+=125$.
  Contours indicate the streamwise vorticity normalized by $u_\tau^*$ and $\nu$.
  $(a)$ $St^+=0.5$; $(b)$ $St^+=5$; $(c)$ $St^+=35$; $(d)$ $St^+=125$; $(e)$ $St^+=0.0$.
   }
  \label{figure10}
\end{figure}

\begin{figure}[t]
     \includegraphics[angle=0, width=\textwidth]{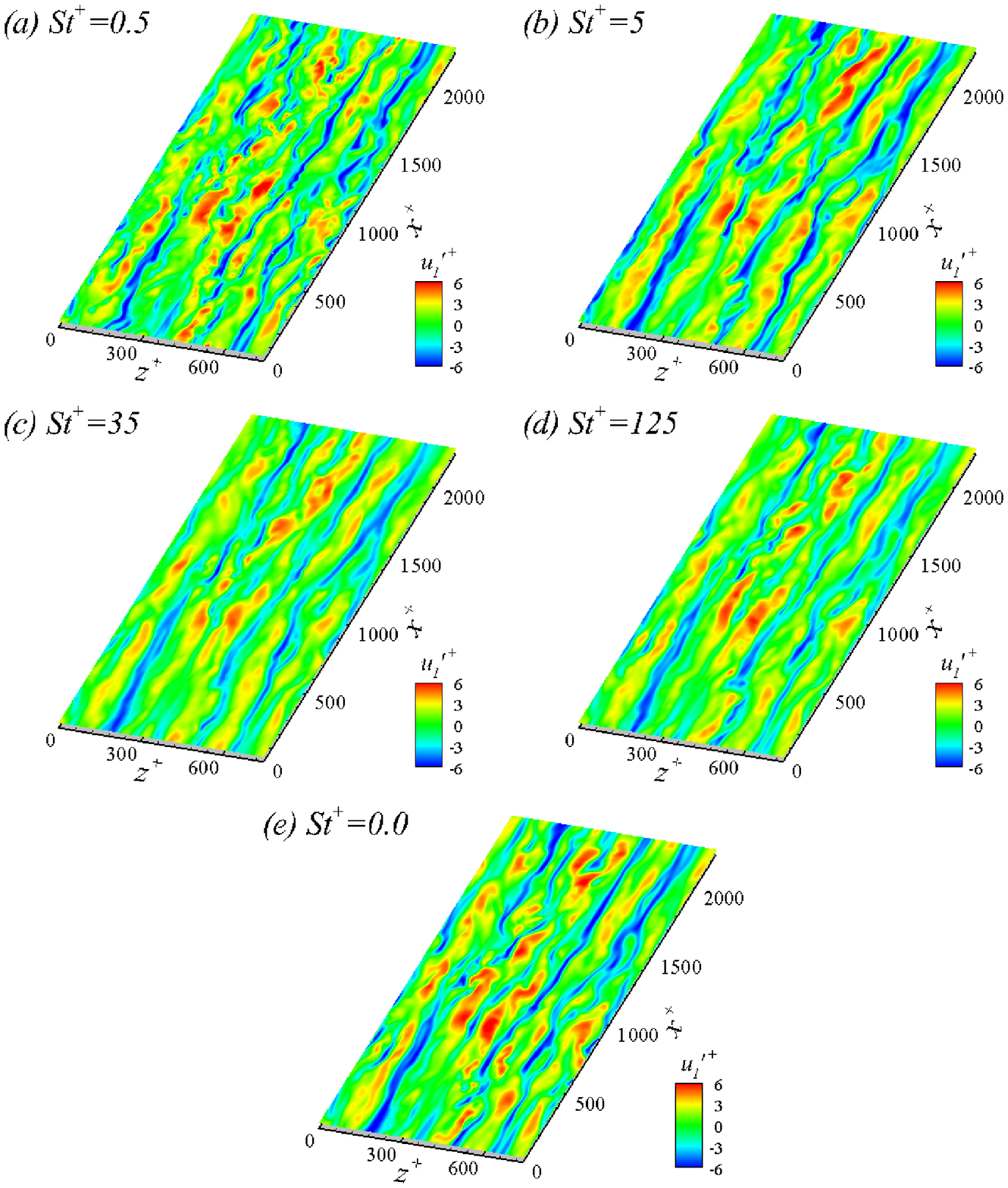}
  \caption{(Color-online) Instantaneous contours of streamwise velocity fluctuation normalized by $u_\tau^*$ 
  at $y^+\approx15$ in the coordinates $(x^+,y^+,z^+)$ at $t^+=125$.
  $(a)$ $St^+=0.5$; $(b)$ $St^+=5$; $(c)$ $St^+=35$; $(d)$ $St^+=125$; $(e)$ $St^+=0.0$. 
  }
  \label{figure11}
\end{figure}

\subsection{Turbulence modification mechanisms}

In the previous sections, 
we presented varying modification of the turbulent statistics and behavior of 
near-wall turbulence structures 
depending on the Stokes number. 
Our aim here is to reveal the effect of Stokes number on 
the physical mechanisms by which particles modify near-wall turbulence structures.
 
\subsubsection{Modification by particles with $\mathbf{St^+=0.5}$} 

\begin{figure}[t]
     \includegraphics[angle=0, width=0.7\textwidth]{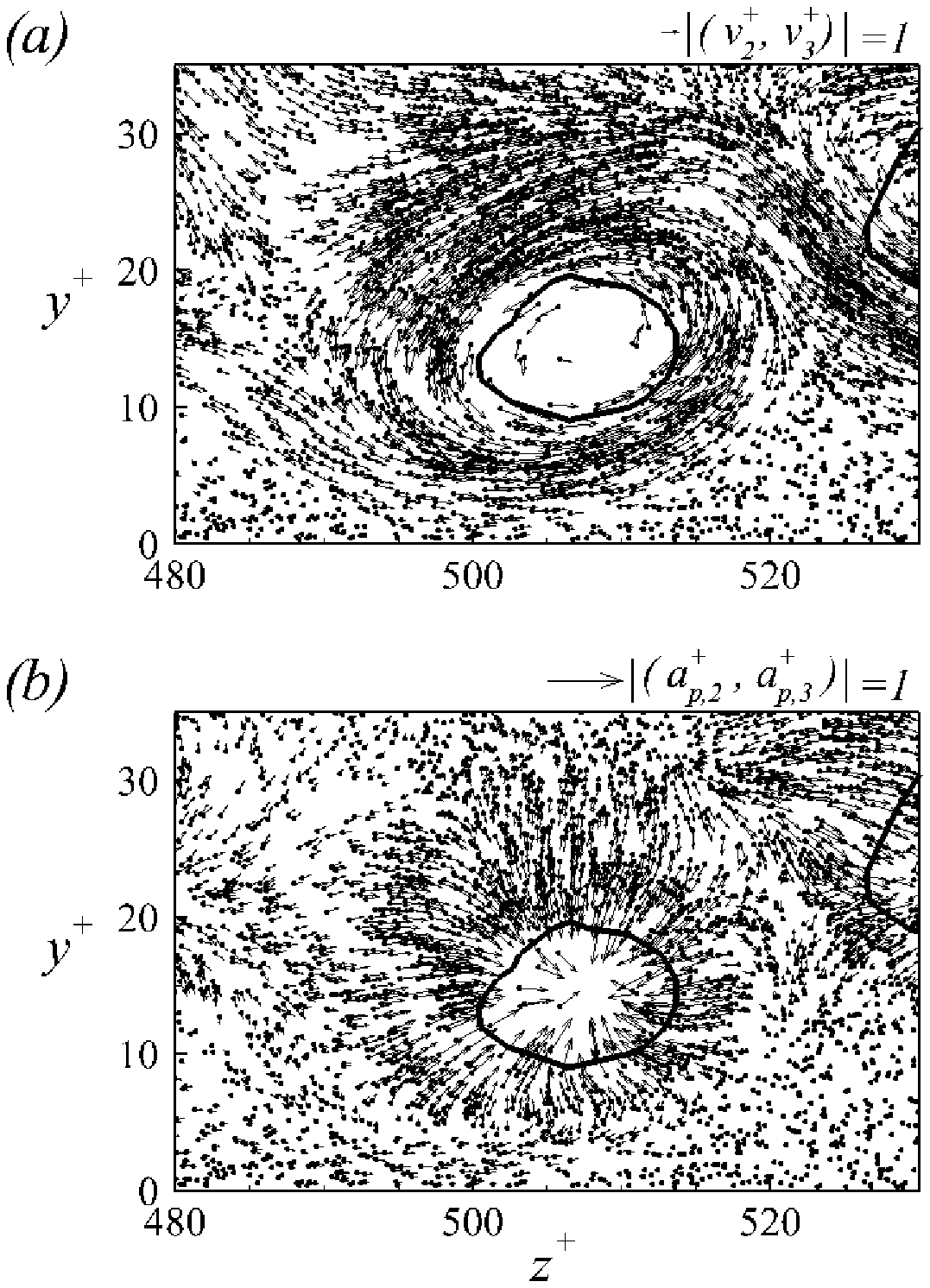}
  \caption{Distributions of particles whose $St^+=0.5$ 
  around a quasistreamwise vortex.
  Dot symbols indicate particle locations. 
  Thick solid lines visualize the edge of a quasistreamwise vortex.
  In $(a)$ and $(b)$, arrows denote  
  particle velocity $(v_2^+,v_3^+)$ and particle acceleration vectors $(a_{p,2}^+,a_{p,3}^+)$ in the $(y^+,z^+)$ plane, respectively. 
  }
  \label{figure12}
\end{figure}

Particles with $St^+=0.5$ have a tendency to follow the fluid particle
since the particle response time is smaller than the Kolmogorov time scale of the fluid, i.e. $St_K<1$ (see Table \rom{1}),
but they have small but finite inertia at the same time. 
This causes the particles trapped in a quasistreamwise vortex to eventually be centrifuged, as shown in Fig. 12.
While being ejected from the quasistreamwise vortex,  
the particles are subject to inward drag in the $(y,z)$ plane, i.e. converging towards the quasistreamwise vortex core (Fig. 12$b$).
Then, the feedback forces, which diverge from the vortex core, cannot directly affect the flow field due to the incompressibility of the fluid, except for creating negative pressure at the core.

Figures 13$(b,c)$ and 13$(d-f)$ show
the respective distributions of the particle feedback force fluctuations, $f_{i}'^+$,
and the local energy exchange between the fluid and particles in the streamwise, wall-normal and spanwise directions, 
i.e. $u_{1}'^+f_{1}'^+$, $u_{2}'^+f_{2}'^+$ and $u_{3}'^+f_{3}'^+$,
around a quasistreamwise vortex in the $(y^+,z^+)$ plane.
Sweep and ejection events generated by the quasistreamwise vortex are also shown in Fig. 13$(a)$.
In Figs. 13$(b,c)$, $f_{i}'^+$ much larger in magnitude than the $\langle f_{i}'^+\rangle$ shown in Fig. 3 
are found around the quasistreamwise vortex. 
Therefore, these large $f_{i}'^+$ approximate $f_i^+$.
It is shown in Fig. 13$(c)$ that 
$(f_{2}'^+,f_{3}'^+)$ of large magnitude, found mostly around the edge of the quasistreamwise vortex, 
diverges from the vortex core region in reaction to the drag forces shown in Fig. 12$(b)$.
Around the quasistreamwise vortex, 
the particle-fluid energy exchange in the wall-normal and spanwise directions, 
i.e. $u_{2}'^+f_{2}'^+$ and $u_{3}'^+f_{3}'^+$,
can be positive or negative, as shown in Figs. 13$(e,f)$,
indicating that particles centrifuged out 
can enhance or suppress the fluid vorticity associated with the quasistreamwise vortex.
This is dependent on the angle between the directions of $(u_{2}'^+,u_{3}'^+)$ and $(f_{2}'^+,f_{3}'^+)$.

\begin{figure}[t]
     \includegraphics[angle=0, width=\textwidth]{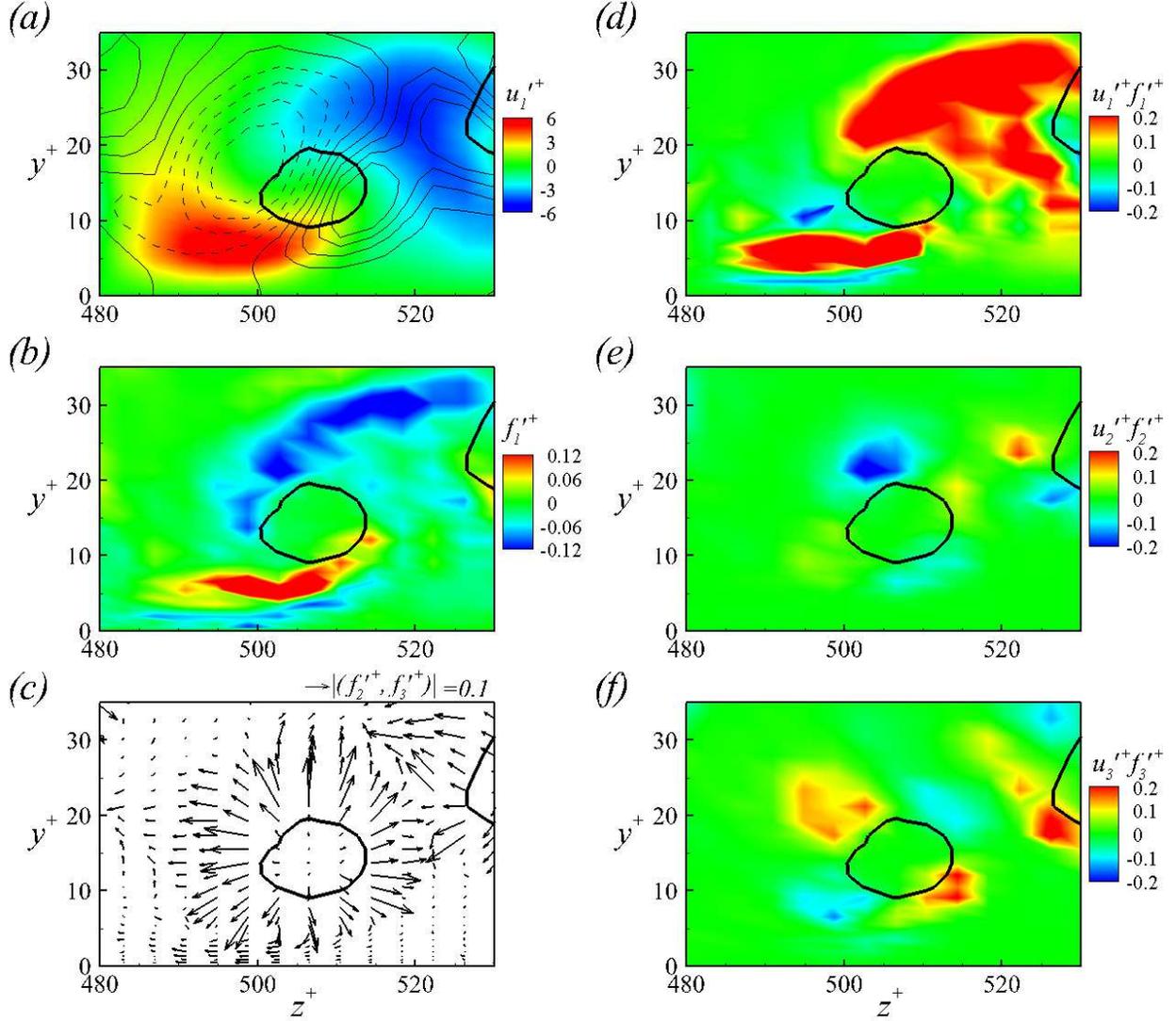}
  \caption{(Color-online) Feedback of particles for which $St^+=0.5$ in the fluid
  around a quasistreamwise vortex in the $(y^+,z^+)$ plane.  
  Thick solid lines illustrate the edge of a quasistreamwise vortex.  
  The vortex is the same as in Fig. 12.
  $(a)$ Sweep ($u_{1}'>0,u_{2}'<0$) and ejection events ($u_{1}'<0,u_{2}'>0$).
  Thin solid and dotted lines represent the positive and negative values of $u_{2}'^+$, respectively.   
  Distributions of $(b,c)$ particle feedback force fluctuations $f_{i}'^+$ and
  $(d,e,f)$ energy exchanges between the fluid and particles $u_{1}'^+f_{1}'^+$, $u_{2}'^+f_{2}'^+$ and $u_{3}'^+f_{3}'^+$.
  In $(c)$, arrows denote particle feedback force fluctuation vectors $(f_{2}'^+,f_{3}'^+)$ in the $(y^+,z^+)$ plane.  
  }
  \label{figure13}
\end{figure}

\begin{figure}[t]
     \includegraphics[angle=0, width=\textwidth]{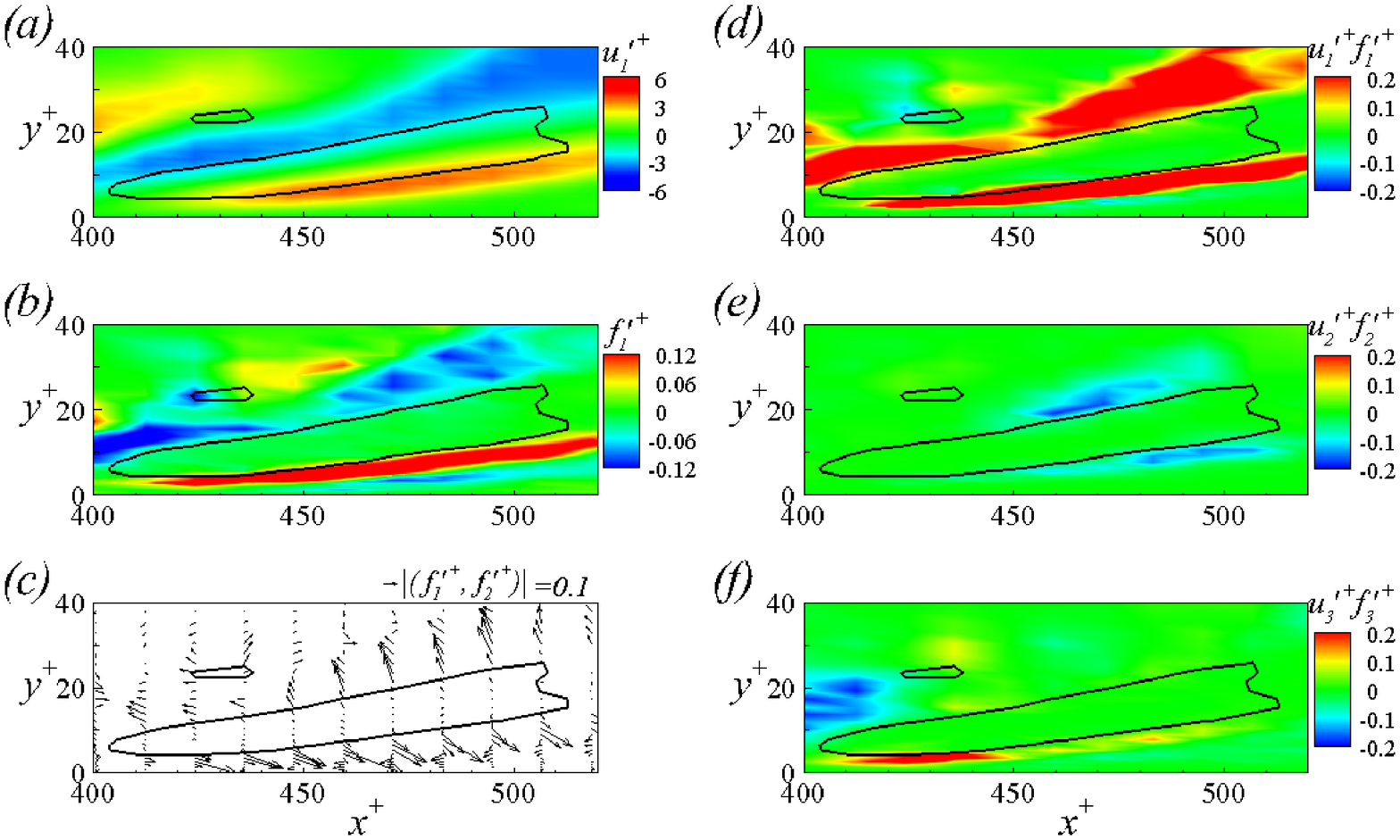}
  \caption{(Color-online) Feedback of particles for which $St^+=0.5$ in the fluid
  around a quasistreamwise vortex, as in Fig. 13, but in the $(x^+,y^+)$ plane.  
  The vortex is the same as in Figs. 12 and 13. 
  }
  \label{figure14}
\end{figure}

From comparison of the results shown in Figs. 13$(a)$ and 13$(b)$, 
it is clear that, around the quasistreamwise vortex,
particles with $St^+=0.5$ 
exert a positive streamwise feedback force 
on the fluid in the sweep region,
and a negative one in the ejection region. 
A plausible explanation for this can be developed by
considering the action of both particle inertia and mean shear near the wall.
When a particle has a small Stokes number, it almost follows the fluid particle, 
but a slip between the two phases eventually occurs due to the small but finite inertia of the particle.
If this particle
is located on the sweep side of a quasi-streamwise vortex,
it migrates 
towards the wall, and leads the fluid due to both particle inertia and shear, 
thereby 
producing positive $f_{1}'^+$ in the sweep region.
On the other hand, 
particles on the ejection side move away from the wall,
and lag the fluid, 
thus producing negative $f_{1}'^+$ in the ejection region. 
In both the sweep and ejection regions,
$u_{1}'^+f_{1}'^+$ becomes positive, as shown in Fig. 13$(d)$,
due to the fact that $f_{1}'^+$ has the same sign as $u_{1}'^+$ in both regions (Fig. 13$b$),
indicating that 
energy is transferred from the particles to the streamwise fluid velocity fluctuations.
Considering that ejections are responsible for the presence of low-speed streaks,
this feedback may cause the low-speed streaks to become strongly unstable
and ultimately influence the birth of new quasistreamwise vortices,
since the proposed mechanisms for vortex formation in near-wall turbulence 
generally involve instability of the streaks.\cite{Robinson91a,Hamilton95,Jimenez99,Schoppa02}
Our results in Figs. 7 and 10$(a,e)$ 
show the increase in the number of quasistreamwise vortices due to particles with $St^+=0.5$ 
compared to the particle-free case,
consistent with the above argument.

Figure 14 shows the feedback of particles
around the quasistreamwise vortex, but in the $(x^+,y^+)$ plane.  
Features of the interactions between the particles and fluid shown in Fig. 13
are clearly seen in the $(x^+,y^+)$ plane.
In Figs. 14$(b,c)$, large $f_{i}'^+$ are found around the quasistreamwise vortex.
In particular, these large $f_{i}'^+$ influence the fluid  
mainly in sweeps and ejections around the quasistremawise vortex, as shown in Fig. 14$(d)$.
In those regions, the energy is transferred from the particles to the fluid in the streamwise direction, 
i.e. $u_{1}'^+f_{1}'^+$ is positive. 
On the other hand, $u_{2}'^+f_{2}'^+$ and $u_{3}'^+f_{3}'^+$ can be both positive and negative
around the quasistreamwise vortices (Figs. 14$e,f$).  

The energy exchange between the fluid and particles for which $St^+=0.5$ in the bottom half of the channel are visualized through three-dimensional isosurfaces in Figs. 15 and 16.
In Fig. 15$(a)$, the plotted $u_{1}'^+f_{1}'^+$ isosurfaces are generally positive around quasistreamwise vortices, 
confirming the results of Figs. 13$(d)$ and 14$(d)$.
In Fig. 15$(b)$, 
the positive $u_1'^+f_1'^+$ isosurfaces above $y^+\approx20$ correlate well with the regions of low-speed streaks,
supporting our argument that the particles enhance the instability of low-speed streaks,
and thus increase the number of quasistreamwise vortices.
The negative $u_1'^+f_1'^+$ isosurfaces are sparsely observed due to the presence of spanwise structures, such as the head of the hairpin-type vortex. 
While $u_1'^+f_1'^+$ is almost entirely positive (Fig. 15$a$), 
the plotted $u_2'^+f_2'^+$ and $u_3'^+f_3'^+$ have both positive and negative values surrounding quasistreamwise vortices, as shown in Fig. 16. 
Although not shown here, both $\langle u_2'^+f_2'^+\rangle$ and $\langle u_3'^+f_3'^+ \rangle$ are negative throughout the channel width with negative peaks in the buffer layer.
This indicates that, unlike in the streamwise case, in the wall-normal and spanwise directions, the particles, on average, act to extract kinetic energy from the fluid around quasistreamwise vortices.
In the period of simulation, however, the streamwise energy transfer appears to have a more significant effect on turbulence modification by the particles.


\begin{figure}[t]
     \includegraphics[angle=0, width=0.7\textwidth]{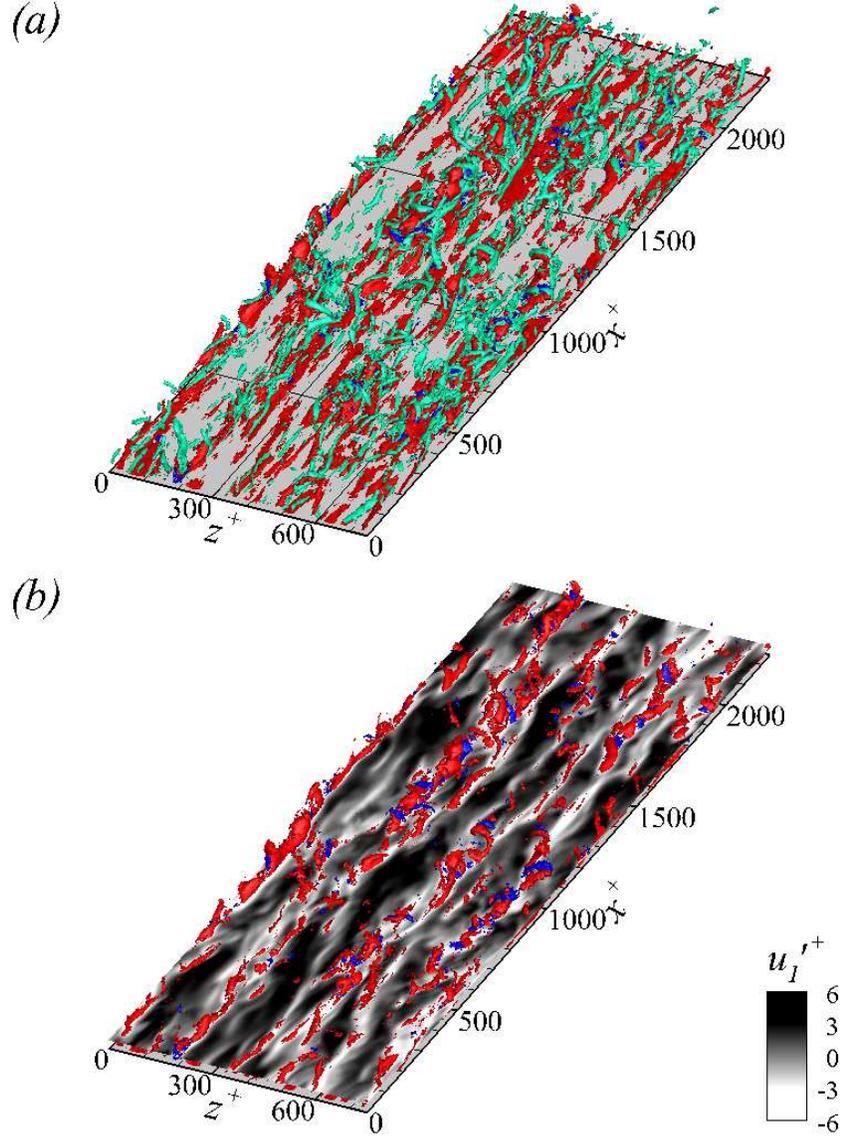}
  \caption{(Color-online) Three-dimensional isosurfaces of $u_{1}'^+f_{1}'^+=0.16$ (red color) and $-0.16$ (blue color) for the case of $St^+=0.5$.
  The value of $0.16$ was chosen to be comparable to the maximum r.m.s. value of $u_{1}'^+f_{1}'^+$ at  the buffer layer.
  $(a)$ Green isosurfaces indicate vortical structures characterized by $\lambda_2^+=-0.02$.
  $(b)$ Gray colors on the $(x^+,z^+)$ plane indicate $u_{1}'^+$ at $y^+\approx20$.
  }
  \label{figure15}
\end{figure}

\begin{figure}[t]
     \includegraphics[angle=0, width=0.7\textwidth]{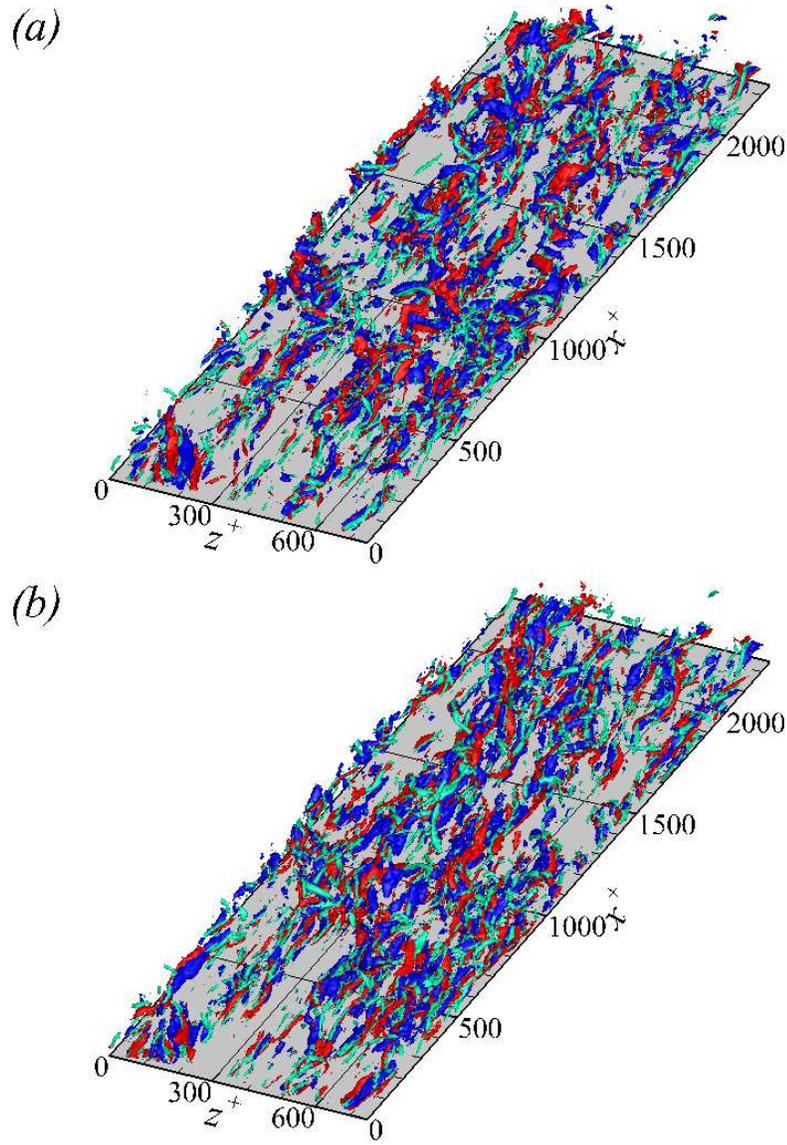}
  \caption{(Color-online) Three-dimensional isosurfaces of $(a)$ $u_{2}'^+f_{2}'^+=0.035$ (red color) and $-0.035$ (blue color)
  and $(b)$ $u_{3}'^+f_{3}'^+=0.055$ (red color) and $-0.055$ (blue color) for $St^+=0.5$.
  The values of 0.035 and 0.055 were chosen to be comparable to the maximum r.m.s. values of $u_{2}'^+f_{2}'^+$ and $u_{3}'^+f_{3}'^+$ at the buffer layer, respectively.
  Green isosurfaces indicate vortical structures characterized by $\lambda_2^+=-0.02$.
  }
  \label{figure16}
\end{figure}

For a small Stokes number such that $St_K\ll1$,
an approximation for particle velocity can be given as\cite{Maxey87,Fouxon12}
\begin{equation}
v_i\approx\tilde{u}_i-\tau_p\frac{Du_i}{Dt}\Bigg|_{\vec{q}(t)}.
\end{equation}
Using Eq. (4), rearrangement of Eq. (11) gives the following relationship 
\begin{equation}
\frac{Du_i}{Dt}\Bigg|_{\vec{q}(t)}\approx\frac{\tilde{u}_i-v_i}{\tau_p}=\frac{dv_i}{dt}.
\end{equation}
Note that in Eqs. (11) and (12), the nonlinear drag correction factor $\gamma$ is not taken into account (i.e. $\gamma=1$),
since $Re_p$ is generally much smaller than 1 when $St_K\ll1$.
Here, it is possible to redefine the particle feedback force for a small Stokes number using Eq. (12) as
\begin{equation}
f_i=-\frac{m_p}{m_f}\sum_{k=1}^{N_p}\left(\frac{dv_i}{dt}\right)_k
\approx-\frac{m_p}{m_f}\sum_{k=1}^{N_p}\left(\frac{Du_i}{Dt}\Bigg|_{\vec{q}(t)}\right)_k.
\end{equation}
In Eq. (13), if the volume containing the $N_p$ particles is sufficiently small,
we can assume that
the ensemble average of fluid acceleration at the locations of the particles $\vec{q}(t)$ 
in the small volume, including point $(x,y,z)$ at time $t$,
$\frac{1}{N_p}\sum_{k=1}^{N_p}(\frac{Du_i}{Dt}|_{\vec{q}(t)})_k$, 
becomes $\frac{Du_i}{Dt}(x,y,z,t)$.
Then, we find a new equation for the particle feedback force field for a small Stokes number as
\begin{equation}
f_i=-\frac{\rho_p}{\rho}\frac{\pi{d}_p^3}{6}n\frac{Du_i}{Dt}.
\end{equation}
As shown in the study by Yeo \emph{et al.}\cite{Yeo10} and in the results of Figs. 8 and 9, 
the r.m.s. values of $\frac{Du_i}{Dt}$ are much higher, at least two times greater than the mean values.
Considering the relationship between $f_i$ and $\frac{Du_i}{Dt}$ in Eq. (14),
this contributes to the higher r.m.s. feedback forces observed compared to their mean values in Fig. 3.
Intermittent events of $(\frac{Du_2}{Dt},\frac{Du_3}{Dt})$ are mostly the centripetal accelerations associated with quasistreamwise vortices,
and large negative $\frac{Du_1}{Dt}$ occurs in sweeps, where high-momentum fluid moves towards the viscous sublayer.\cite{Lee04,Yeo10}
Furthermore, a close investigation reveals that, in the ejection regions where low-momentum fluid is lifted away from the wall, 
positive $\frac{Du_1}{Dt}$ occurs.
Since the feedback force $f_i$ expressed in Eq. (14) 
is opposite to the direction of fluid acceleration $\frac{Du_i}{Dt}$,
Eq. (14) describes the distributions of $f_{i}'$ of large magnitude well in Figs. 13$(b,c)$ and 14$(b,c)$,
where large $(f_{2}',f_{3}')$ is directed away from the vortex core and large positive and negative $f_{1}'$ are observed in sweeps and ejections, respectively.
Equation (14) also explains that 
the magnitude of the particle feedback force increases in proportion to the particle-to-fluid density ratio, particle diameter (i.e. its volume) and particle number density. 

Combining Eqs. (1) and (14),
the modified momentum equation for the fluid 
due to the presence of particles whose Stokes number is small enough
can be obtained as
\begin{equation}
\left(1+\frac{\rho_p}{\rho}\frac{\pi{d}_p^3}{6}{n}\right)\frac{Du_{i}}{Dt}
=-\frac{1}{\rho}\frac{\partial{p}}{\partial{x_i}}
+\nu\frac{{\partial}^2{u_i}}{\partial{x_j}\partial{x_j}}.
\end{equation}
Equation (15) clearly shows that when Stokes number is sufficiently small,
particles enhance the fluid density by a factor of $1+\frac{\rho_p}{\rho}\frac{\pi{d}_p^3}{6}{n}$.
It should be noted that $\frac{\rho_p}{\rho}\frac{\pi{d}_p^3}{6}{n}$ is the local mass loading.
Given that the horizontally averaged near-wall number density, $\overline{n}$, is roughly the same order as the initial number density ($\overline{n}_0=0.3$), as shown in Fig. 1, the local density is equivalently increased by 30\% on average.
The resultant effect is an increase of fluid inertia and, thus, destabilization.
This argument supports our numerical results for $St^+=0.5$,
in that the simulated particles transfer their energy to the fluid in the high-speed regions and low-speed streaks,
exerting their reaction forces on streamwise fluid velocity fluctuations in the same direction. 
This argument is also consistent with the previous analysis of Saffman\cite{Saffman62} 
who considered the destabilization of a dusty gas laminar flow
and the numerical results for the mixing layer,\cite{Tong99} decaying isotropic turbulence\cite{Druzhinin99} and jet.\cite{DeSpirito01}
This also explains why previous simulation studies at similar Stokes numbers\cite{Pan96, Zhao11} did not report
the enhancement of turbulence by particles. In Pan and Banerjee,\cite{Pan96} gravity settled particles towards the wall so quickly that the resulting mass loading around quasistreamwise vortices was too low to influence the flow, while in Zhao and Andersson\cite{Zhao11}, the initial mass loading was too low.
Furthermore, in Rashidi \emph{et al.},\cite{Rashidi90} the settling particles may have weakened the streamwise energy transfer responsible for the enhancement of turbulence shown above.

\subsubsection{Modification by particles with $\mathbf{St^+=}$ 5, 35 and 125} 

In the previous section, 
we showed the role of particles with $St^+=0.5$ 
in the modification of turbulence in the near-wall region.
When the Stokes number is large,  
particles modify turbulent channel flow differently than when Stokes number is small.
Dritselis and Vlachos\cite{Dritselis08,Dritselis11} 
showed the attenuation of fluid velocity and vorticity associated with quasistreamwise vortices 
in particle-laden turbulent channel flows 
for $St^+=10, 25, 100$ and 200 
by investigating conditionally averaged results.
They explained that due to inertia particles are dragged by the fluid, 
thus simultaneously producing feedback forces to oppose fluid velocities. 
These feedback forces act as a torque on quasistreamwise vortices in the opposite direction of the rotation of the vortices.
Also, particles suppress streamwise fluid velocity fluctuations, particularly in low-speed streaks.
Consequently, quasistreamwise vortices were depopulated by particles.
Furthermore, Dritselis and Vlachos showed that the effects of particles on the fluid flows are enhanced for small Stokes numbers ($St^+=10$ or 25) due to preferential concentration.
Here, we examine how particles physically modify turbulence in the near-wall region when $St^+=5, 35$ and 125
through instantaneous three-dimensional isosurfaces of the energy exchange between the fluid and particles.

The most significant attenuation of the turbulence occurs in the case of $St^+=35$, as shown in Sec. \rom{3} A and Sec. \rom{3} B.
In this case, a negative contribution of $f_{1}'^+$ to $u_{1}'^+$, i.e. $u_{1}'^+f_{1}'^+<0$, becomes significant
compared to the case of $St^+=0.5$ (Figs. 17$b$ and 18$b$).  
Furthermore, 
in Fig. 18$(b)$, the regions with negative $u_{1}'^+f_{1}'^+$
nearly coincide with the regions of low-speed streaks, forming a streaky pattern.
A possible reason is that the particles that populate the outer layer 
preferentially collect in the low-speed streaks near the wall
while maintaining their higher momentum.
Therefore, particles have positive $f_{1}'$ in the low-speed streaks characterized by regions of $u_{1}'<0$, and thus produce negative $u_{1}'f_{1}'$. 
This is opposite to the case of $St^+=0.5$.
In Figs. 19$(b)$ and 20$(b)$, it is apparent that 
the plotted $u_{2}'^+f_{2}'^+$ and $u_{3}'^+f_{3}'^+$ for the case of $St^+=35$  
are almost entirely negative.
Furthermore, regions with negative $u_{2}'^+f_{2}'^+$ and $u_{3}'^+f_{3}'^+$ are localized around the quasistreamwise vortices,
since particles with $St^+=35$ 
are primarily transported along curved streamlines around quasistreamwise vortices,\cite{Marchioli02}
and their slow responses to fluid velocities generate $(f_{2}',f_{3}')$ against $(u_{2}',u_{3}')$. 
Therefore, the particles obstruct the wall-normal and spanwise fluid motions around quasistreamwise vortices,
reducing the fluid vorticty associated with the vortices.

In the case of $St^+=125$,
due to their large inertia, 
the particles are less affected by quasistreamwise vortices.
In this case, preferential concentration rather decreases compared to the case of $St^+=35$, as shown in Fig. 1.
Therefore, the correlation between regions with negative $u_{1}'^+f_{1}'^+$ and $u_{1}'<0$ decreases 
(Figs. 17$c$ and 18$c$)
and the negative isosurfaces of $u_{2}'^+f_{2}'^+$ and $u_{3}'^+f_{3}'^+$ are more dispersed 
than in the case of $St^+=35$ (Figs. 19$c$ and 20$c$).
As a result, the degree of turbulence attenuation is reduced compared to the case of $St^+=35$.

For $St^+=35$ and 125, our discussion based on instantaneous three-dimensional snapshots 
confirms the conditionally averaged results of Dritselis and Vlachos.\cite{Dritselis08,Dritselis11}
On the other hand, when $St^+=5$,
Figs. 17$(a)$, 18$(a)$, 19$(a)$ and 20$(a)$ demonstrate that 
the effects of particles with $St^+=0.5$ and $St^+=35$ are present together at the same time.
In this case, both the positive and negative contributions of $f_{1}'$ to $u_{1}'$ are found 
in low-speed streaks. 
Furthermore, the negative $u_2'^+f_2'^+$ and $u_3'^+f_3'^+$ are more common than the positive ones around quasistreamwise vortices.
The net effect is to attenuate the turbulence, but with less influence than in the case of $St^+=35$.

\begin{figure}
     \includegraphics[angle=0, width=0.65\textwidth]{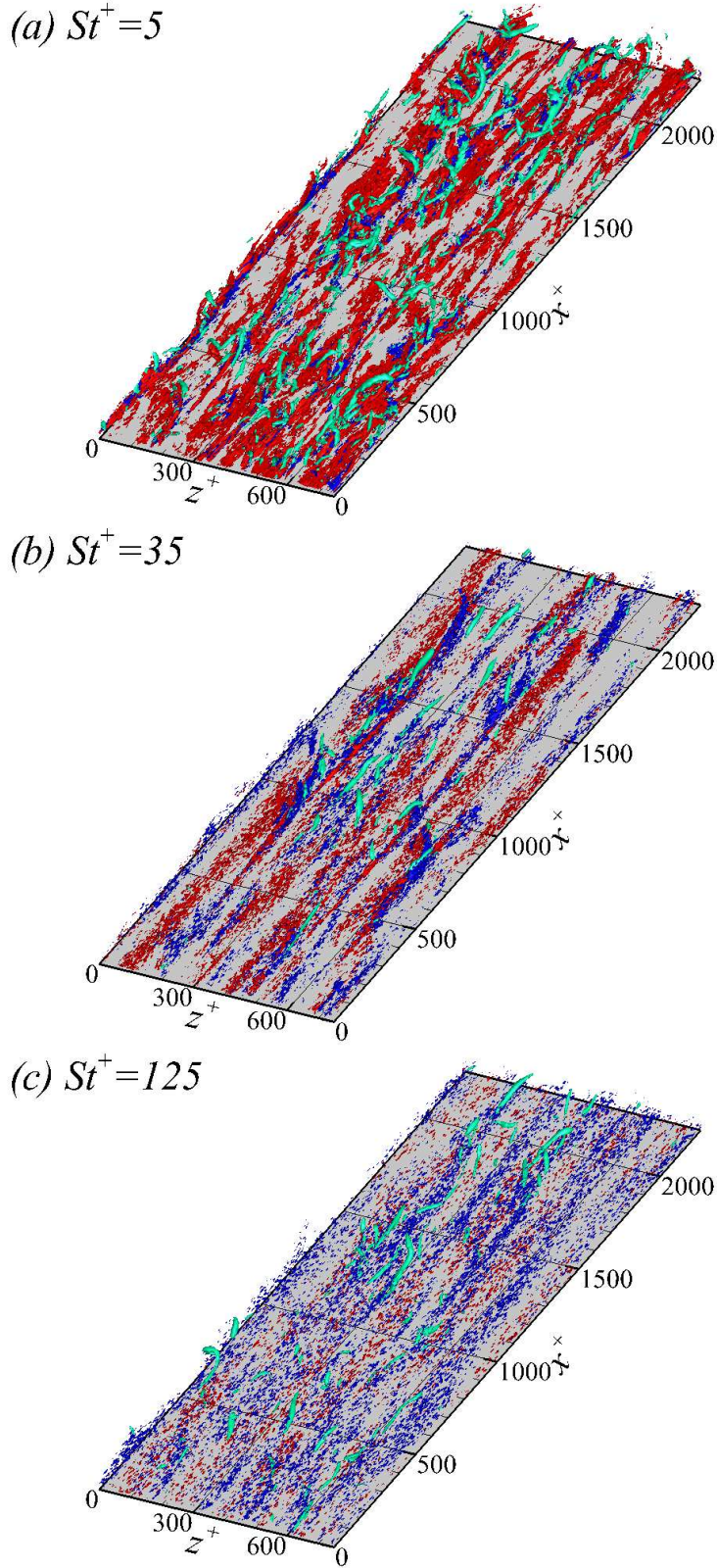}
  \caption{(Color-online) Three-dimensional isosurfaces of $u_{1}'^+f_{1}'^+=0.16$ (red color) and $-0.16$ (blue color).
  Green isosurfaces indicate vortical structures characterized by $\lambda_2^+=-0.02$.
  $(a)$ $St^+=5$; $(b)$ $St^+=35$; $(c)$ $St^+=125$.
  }
  \label{figure17}
\end{figure}

\begin{figure}
     \includegraphics[angle=0, width=0.65\textwidth]{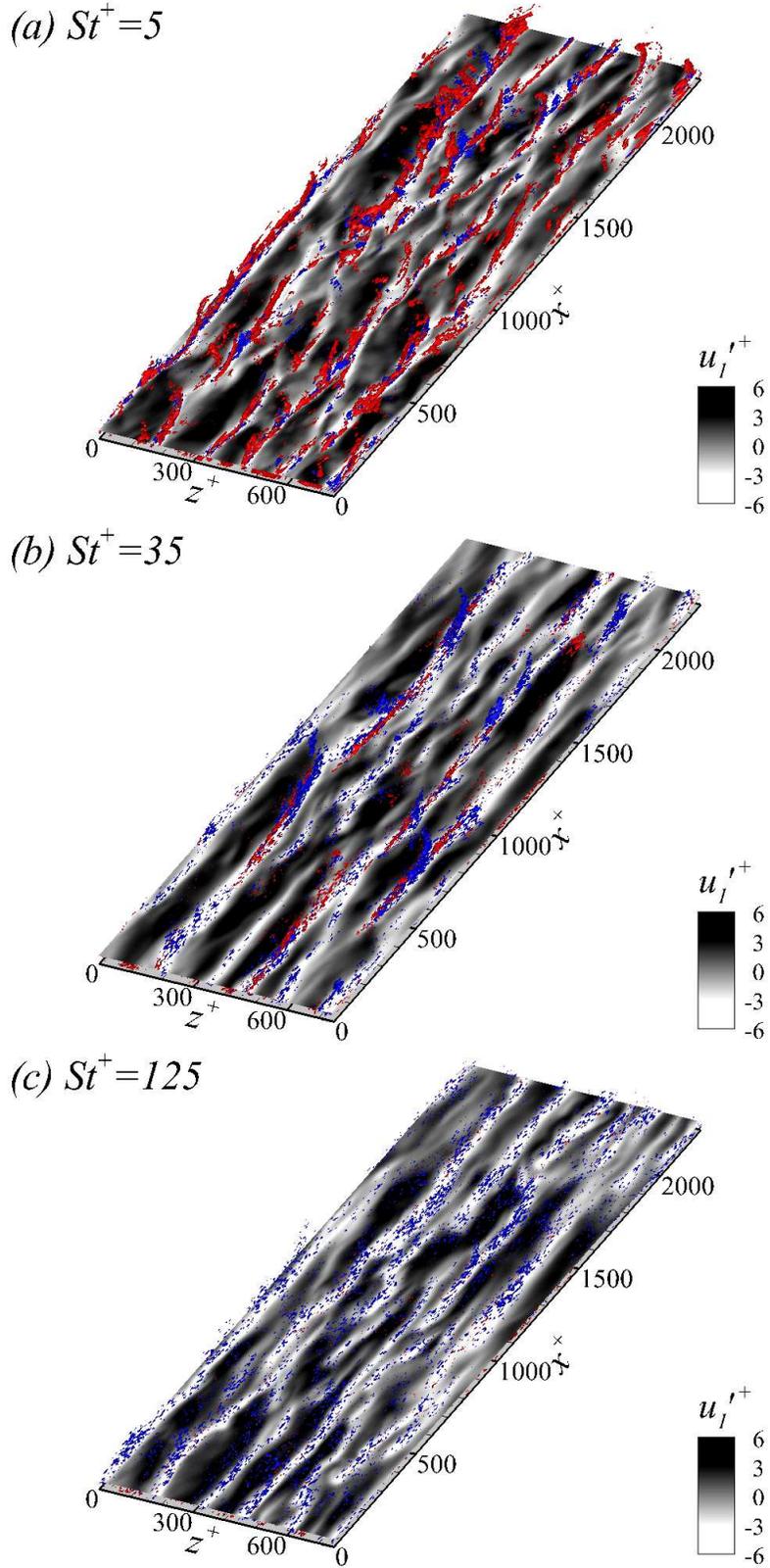}
  \caption{(Color-online) Three-dimensional isosurfaces of $u_{1}'^+f_{1}'^+=0.16$ (red color) and $-0.16$ (blue color).
  Gray colors on the $(x^+,z^+)$ plane indicate $u_{1}'^+$ at $y^+\approx20$.
  $(a)$ $St^+=5$; $(b)$ $St^+=35$; $(c)$ $St^+=125$.
  }
  \label{figure18}
\end{figure}

\begin{figure}
     \includegraphics[angle=0, width=0.65\textwidth]{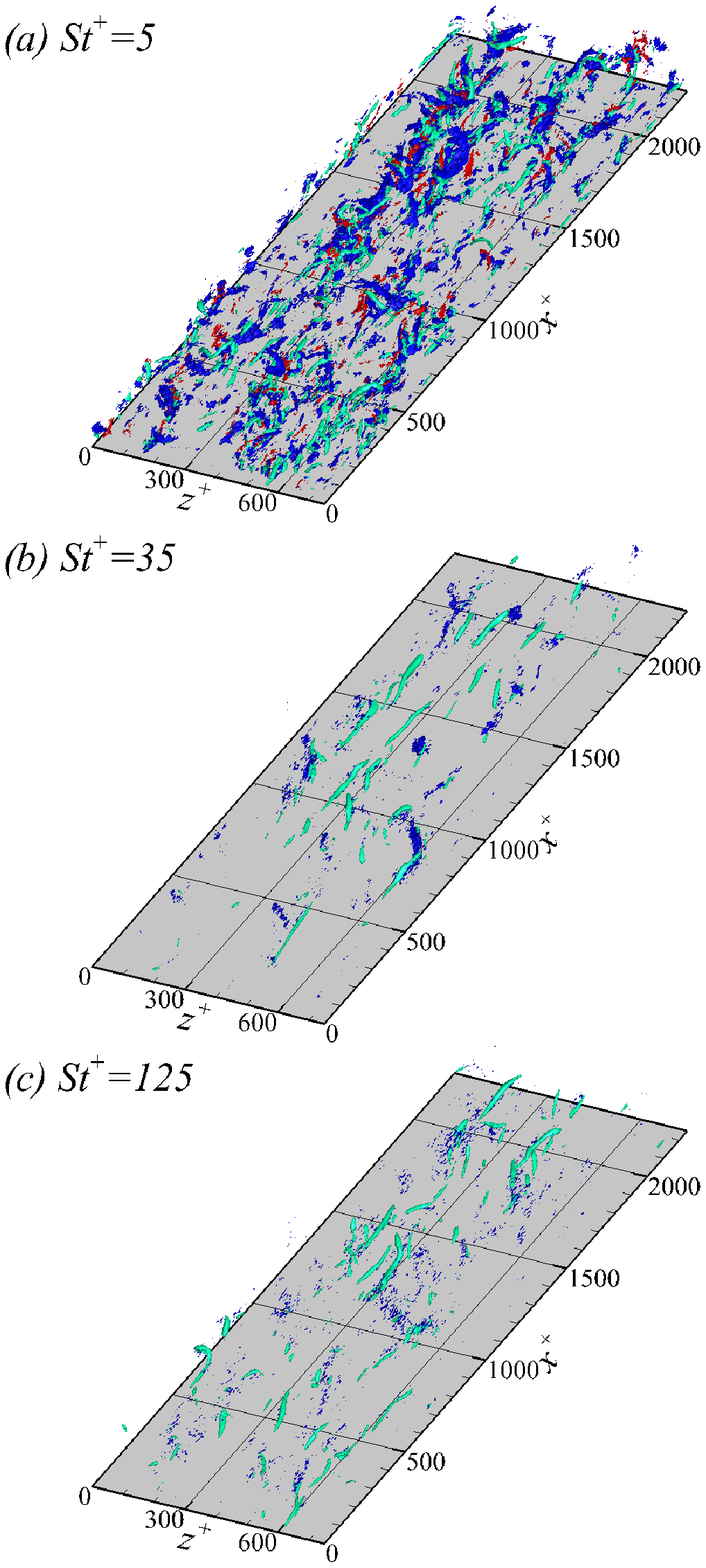}
  \caption{(Color-online) 
  Three-dimensional isosurfaces of $u_{2}'^+f_{2}'^+=0.035$ (red color) and $-0.035$ (blue color).
  Green isosurfaces indicate vortical structures characterized by $\lambda_2^+=-0.02$.
  $(a)$ $St^+=5$; $(b)$ $St^+=35$; $(c)$ $St^+=125$.
  }
  \label{figure19}
\end{figure}

\begin{figure}
     \includegraphics[angle=0, width=0.65\textwidth]{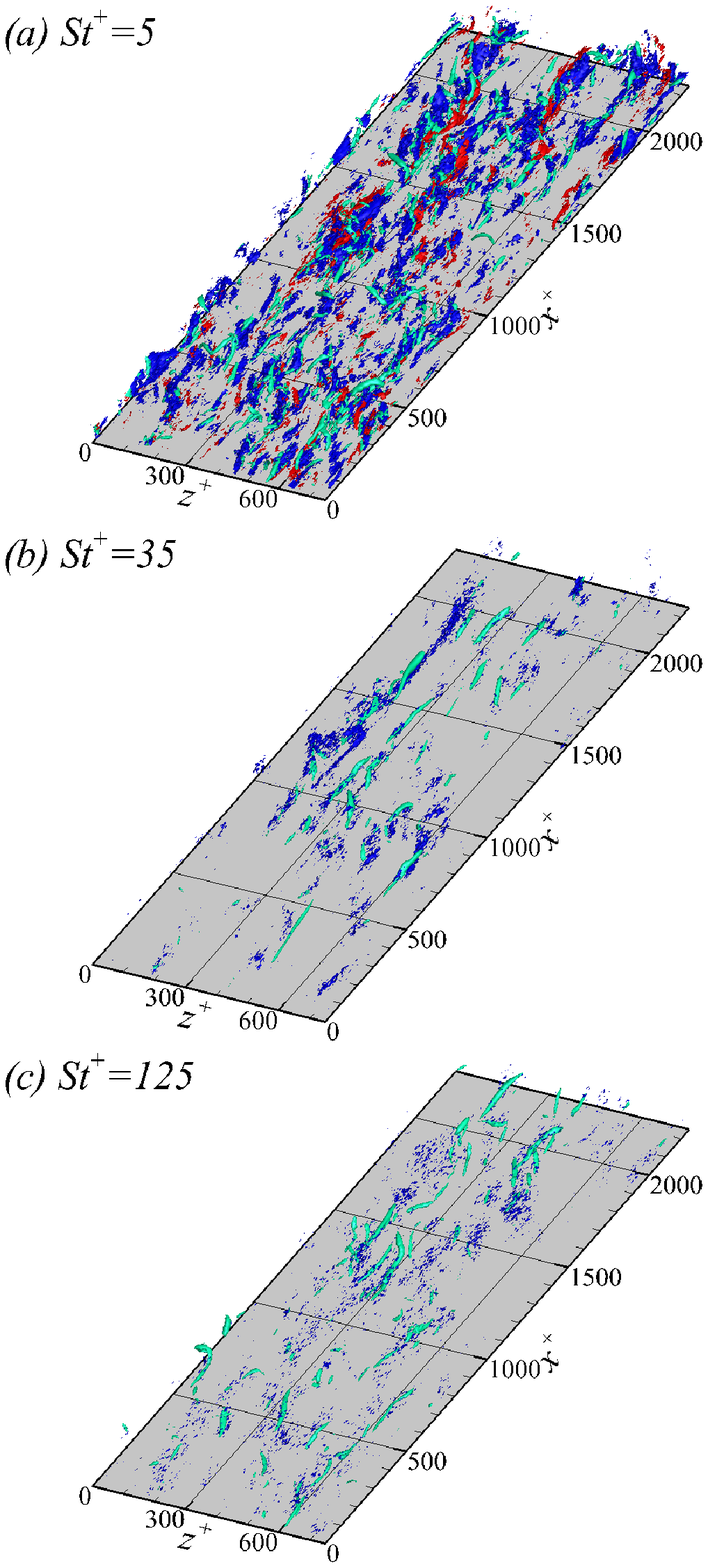}
  \caption{(Color-online) Three-dimensional isosurfaces of $u_{3}'^+f_{3}'^+=0.055$ (red color) and $-0.055$ (blue color).
  Green isosurfaces indicate vortical structures characterized by $\lambda_2^+=-0.02$.
  $(a)$ $St^+=5$; $(b)$ $St^+=35$; $(c)$ $St^+=125$.
  }
  \label{figure20}
\end{figure}

\section{Conclusions}

The effect of Stokes number on turbulence modification  
in a channel was studied using direct numerical simulation
with a point-force approximation
for small, heavy particles with a diameter smaller than the Kolmogorov length scale.
Stokes numbers of $St^+=0.5, 5, 35$ and 125 were chosen to cover a wide range.
Within this range, we find that the modification of turbulent channel flow is not monotonic. 

We find that particles with $St^+=0.5$ increase the number of quasistreamwise vortices,
while particles with larger Stokes numbers decrease it.
The turbulence statistic is augmented by particles with $St^+=0.5$
and suppressed by particles with larger Stokes numbers through the modification of vortex distribution.
This attenuation is most pronounced for the case of $St^+=35$ where preferential concentration is maximized.

The physical mechanism for turbulence augmentation by particles with $St^+=0.5$ can be explained as follows.
Particles with this low inertia almost follow the fluid particles, 
but a slip between the two phases eventually occurs due to their small but finite inertia.
If this particle is located on the sweep side of a quasistreamwise vortex,
it moves towards the wall along the fluid and is decelerated due to the shear,
thereby accelerating the fluid according to the action-reaction principle.
Based on the same principle, the particles decelerate the fluid in ejections. 
Consequently, energy transfer occurs 
from particles to fluid velocity fluctuations in high-speed regions and low-speed streaks.
This feedback on the fluid may increase the instability of low-speed streaks
and ultimately influence the birth of new quasistreamwise vortices.

The most significant attenuation of turbulence is observed for the case of $St^+=35$. 
At this Stokes number, preferential concentration is maximized.
Therefore, particles populating the outer layer 
preferentially collect in low-speed streaks near the wall while maintaining their original higher momentum,
and lead the fluid there, thereby accelerating the low-speed fluid. 
This is opposite to the case of the smallest Stokes number.
Furthermore, the slow responses of particles to the fluid bring about 
feedback against fluid velocities in the wall-normal and spanwise directions.
In particular, this feedback is localized around quasistreamwise vortices, 
since the particles are transported along preferred paths around the vortices, i.e. in sweeps and ejections.
Therefore, particles decrease the fluid vorticity associated with quasistreamwise vortices.

In the case of $St^+=125$, the particles are less affected by quasistreamwise vortices due to their large inertia, and
their preferential concentration rather decreases compared to particles with $St^+=35$.
In this case, the feedback of the particles is scattered. 
This reduces the degree of turbulence attenuation compared to the case of $St^+=35$.

In the case of $St^+=5$, 
both the effects of case $St^+=35$ as well as case $St^+=0.5$ appear. 
The net result is to damp quasistreamwise vortices, but with less influence than $St^+=35$.
 
Our simulations focus on a relatively early stage of the preferential concentration of particles
in order to highlight the immediate effect of particle motion depending on Stokes number.
For example, after a long period of time, particles with $St^+=0.5$ accumulate at the wall 
due to their small but finite inertia and persistent interaction with near-wall quasistreamwise vortical structures,
and this will mask the pure effect of small Stokes number demonstrated in the present study.

\begin{acknowledgments}
This research was supported by the National Research Foundation of Korea (NRF) funded by the Korean Government (MSIP, ME) (20090093134, 2014R1A2A2A01006544, 2011-0008788). Most computations were carried out at the KISTI Supercomputing Center.
\end{acknowledgments}

\bibliography{ref}

\end{document}